\documentstyle[tighten,multicols,eqsecnum,pre,aps]{revtex}

\begin{document}

\draft

\title{Stability of cylindrical domains in phase-separating binary 
fluids in shear flow} 
\author{Amalie Frischknecht} 
\address{Department of Physics, University of 
California, Santa Barbara, California 93106-4030}
\date{May 15, 1998}

\maketitle

\input epsf

\begin{abstract}

The stability of a long cylindrical domain in a phase-separating 
binary fluid in an external shear flow is investigated by linear 
stability analysis.  Using the coupled Cahn-Hilliard and Stokes 
equations, the stability eigenvalues are derived analytically for long 
wavelength perturbations, for arbitrary viscosity contrast between the 
two phases.  The shear flow is found to suppress and sometimes 
completely stabilize both the hydrodynamic Rayleigh instability and 
the thermodynamic instability of the cylinder against varicose 
perturbations, by mixing with nonaxisymmetric perturbations.  The 
results are consistent with recent observations of a ``string phase'' 
in phase-separating fluids in shear.

\end{abstract}


\begin{multicols}{2}

\section{Introduction}

Phase-separating binary fluids form complex patterns of domains after 
a quench into the two-phase region of the phase diagram.  The domain 
morphology is determined by a number of factors such as the volume 
fractions of the two phases, their viscosities, and any external 
forces applied to the system \cite{siggia,onuki94}.  Of particular 
interest here is the effect of an external shear flow applied to the 
fluid.  The shear flow competes with the phase-separation process, 
influencing the morphology and stability of the domains.  Besides 
being a fascinating problem in nonequilibrium physics, this question 
is of practical significance because the final properties of 
industrial materials involving binary fluids often depend on the 
domain morphology.

At late times after a quench into an unstable state, a 
phase-separating binary fluid consists of domains of the two phases 
which coarsen with time.  The presence of a shear flow dramatically 
alters the kinetics of the phase separation.  The effects of the 
deformation by the shear flow depend on the relations between the 
various time scales in the system.  The characteristic time scale for 
the shear flow is just the inverse shear rate, $1/\dot{\gamma}$.  When 
this time is shorter than the characteristic relaxation time of 
thermal concentration fluctuations $\tau_{\xi}$, $\dot{\gamma} 
\tau_{\xi} > 1$, the system is in a ``strong-shear'' regime in which 
the critical fluctations are modified by the flow.  Conversely, in the 
``weak-shear'' regime $\dot{\gamma} \tau_{\xi} < 1$ the critical 
fluctuations are not affected by the flow.  A third time scale, which 
will be crucial here, is associated with the domains.  Clearly when 
the growth rate of the domains, or the growth rate of any 
instabilities associated with the domains, is smaller than the time 
associated with the flow, the shear flow will affect the morphology 
and stability of the domains.

Of interest in this paper is the competition between the shear flow 
and the coarsening process.  The shear flow tends to deform and 
elongate or fragment domains, whereas the thermodynamics favors 
coarsening to larger, isotropic domains.  This competition leads to 
the formation of a nonequilibrium, dynamic steady state in which the 
coarsening is stopped by the shear flow 
\cite{onuki86,hashimoto88,Gold:Min93}.  When the viscosities of the 
two phases are similar and one phase forms a droplet phase, the steady 
state consists of somewhat deformed droplets of the order of the 
Taylor breakup size $R \sim \sigma/\eta \dot{\gamma}$, where $\sigma$ 
is the surface tension, $\eta$ the viscosity and $\dot{\gamma}$ the 
shear rate \cite{Gold:Min93,taylor,rallison,Gold:Min94}.  On the other 
hand, when the two phases are both percolated the morphology is an 
anisotropic bicontinuous structure with apparently stable domains 
highly elongated along the flow direction.  The anisotropy in these 
bincontinuous patterns is much larger than the aspect ratio of 2 or 3 
seen for isolated droplets \cite{baum}.  Microscope observations have 
shown that the domains can be elongated into long cylinders 
\cite{string95,string96,kim}.  In weak shear, $\dot{\gamma} \tau_{\xi} 
< 1$, these stringlike domains still undergo frequent breakup, 
reconnection, and branching, whereas in the strong-shear regime the 
system forms a ``string phase'' consisting of macroscopically long 
cylindrical domains aligned with the flow direction.  These are 
surprising observations, since a long fluid cylinder at rest is 
unstable against breaking up into spherical droplets via the Rayleigh 
capillary instability \cite{rayleigh,tomotika}.  In the situation 
under consideration here, the string is a domain of one phase immersed 
in the other, which we would also expect to be thermodynamically 
unstable since the cylinder could lower its surface energy by 
spheroidizing.  Thus the shear flow stabilizes both the thermodynamic 
instability towards phase separation and the hydrodynamic instability 
of these highly elongated domains.

The goal of this paper is to explore the stabilization of cylindrical 
domains by an imposed shear flow.  It is a sequel to my previous work 
on the stabilization by shear flow of a two-dimensional, lamellar 
domain in phase-separating binary fluids \cite{frisch1}.  In that 
paper it was shown that a lamellar domain at rest with diffuse 
interfaces is unstable towards a ``varicose'' instability.  This 
instability is essentially a coarsening effect and depends on the 
finite width of the interfaces; in the limit of mathematically sharp 
interfaces a lamellar domain is stable.  An external shear flow 
stabilizes the lamellar domain by advecting the top and bottom 
interfaces with respect to each other so that they no longer can
maintain the exact phase relation which produces the unstable varicose 
mode.

\begin{figure}
\begin{center}
\leavevmode
\epsfbox{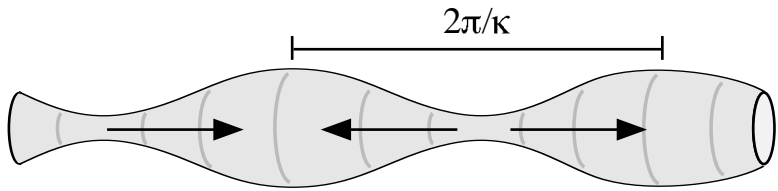}
\end{center}
\caption{\label{varicose} Varicose instability of a fluid cylinder, of
wavelength $2\pi/\kappa$.}
\end{figure}

The instability of a quiescent cylindrical domain of one phase 
immersed in the other is somewhat different, owing to the different 
dimensionality.  There are two separate forces driving instability, 
one hydrodynamic and one thermodynamic.  Rayleigh \cite{rayleigh} and 
later Tomotika \cite{tomotika} analyzed the instability of an 
infinitely long viscous fluid cylinder immersed in an immiscible fluid 
to axisymmetric varicose perturbations as shown in Fig.\ 
\ref{varicose}.  When the wavelength of the varicosity is equal to or 
longer than the circumference of the cylinder the perturbation is 
unstable and grows.  This is because the higher curvature in the necks 
leads to a higher Laplace pressure there than in the bulges, which 
tends to drive fluid from the necks toward the bulges.  Eventually the 
cylinder will break up into spherical droplets, with less total 
surface energy than the original cylinder.  However, even in the 
absence of fluid motion (e.g.\ consider a cylindrical domain in a 
solid binary alloy) a cylindrical domain in the two-phase state is 
still unstable.  This is due to the Gibbs-Thomson effect, in which the 
chemical potential depends on the curvature \cite{frisch1,langer92}.  
The higher curvature at the necks will drive a diffusive flux towards 
the bulges, also leading to instability.  Both of these mechanisms are 
present even in the limit that the interfaces are mathematically 
sharp.

In this paper I perform a linear stability analysis to investigate the 
effect of an external shear flow on the stability of a single 
infinitely long cylindrical domain, perfectly aligned with the flow.  
I consider late times after a quench into the two-phase region, when 
the fluid consists of domains of the two phases close to their 
equilibrium concentrations, separated by well-defined interfaces.  The 
formulation of the problem allows in principle for diffuse interfaces 
with a finite width $\xi$ (when the viscosities of the two phases are 
equal), but in practise results are much easier to obtain in the limit 
of mathematically sharp interfaces and the corrections due to finite 
widths will not affect the results qualitatively.  The shear rate 
$\dot{\gamma}$ is assumed to be small enough that the system is in the 
weak-shear regime, $\dot{\gamma} \tau_{\xi} < 1$, so that the shear 
flow will not influence the structure of the interfaces themselves.  
In this case the usual hydrodynamic equations for a phase-separating 
fluid are valid.  As mentioned above, the string phase itself seems to 
form only in the strong-shear regime \cite{note1}.  However, here I 
will be concerned not with the formation of the string phase but with 
its stability.  The goal is to understand the mechanism by which the 
shear stabilizes these remarkably elongated domains.  The results may 
also illuminate the stability of the highly anisotropic, bicontinuous 
morphologies observed in weak shear.  I will neglect the ends of the 
string and also the possiblity that it could be inclined at a small 
angle to the flow direction.  This approximation seems reasonable 
given the extraordinarily high aspect ratio observed for the strings 
and the fact that long slender drops in shear have a long central 
portion which is cylindrical and aligned with the flow \cite{hinch}.

As well as shedding light on the stability of elongated domains in 
phase-separating fluids, this work encompasses the problem of the 
effect of shear flow on the purely hydrodynamic viscous Rayleigh 
instability in immiscible fluids (neglecting the thermodynamic 
effects).  To my knowledge this problem has not been solved before in 
the particular limit examined here.  Russo and Steen \cite{russo} and 
Lowry and Steen \cite{lowry,lowry2} found that axial flows can 
suppress capillary instabilities on cylindrical interfaces.  Most 
other studies in the fluid dynamics literature concerning the effects 
of externally imposed flows on long fluid cylinders have been in the 
context of drop breakup.  Several authors have considered the linear 
stability of an infinitely long fluid cylinder in an elongational flow 
\cite{tomotika2,mikami,khakhar,janssen}.  The flow field limits the 
growth of any disturbance to a finite value so that there is no true 
instability, and the cylinder is stabilized.  However, some 
disturbances have time to grow transiently to a finite amplitude 
comparable to the decreasing radius of the elongating cylinder, 
causing breakup (even though the disturbances do not grow 
exponentially).  Khakhar and Ottino \cite{khakhar} extended the 
analysis to general linear flows including shear flow, but only in the 
case of small asymmetry, when the shear part of the flow is small 
compared to the stretching.  In this paper I will explore the opposite 
limit in which the stretching is negligible but the asymmetry is 
large.  Finally, Hinch and Acrivos studied a finite, long slender drop 
in shear flow \cite{hinch}.  They find steady-state solutions for the 
shape of the drop at all shear rates, but these equilibrium solutions 
are unstable above a critical shear rate.  This is essentially due to 
the fact that the ends of the drop are not completely aligned with the 
flow, so that at sufficiently high shear rates the drop cannot balance 
the stretching of the ends and it extends transiently, becoming 
progressively thinner.  This does not happen for an infinitely long 
cylinder, as considered in this paper.

In Sec.\ \ref{eqtns} I will describe the model equations of motion 
used to describe the fluid.  In Sec.\ \ref{stab} the equations of 
motion are linearized for small perturbations about a cylindrical 
domain.  Approximate solutions can be found by writing the stability 
eigenvalue equation as a matrix equation in a truncated set of 
``basis'' states, corresponding to different perturbation modes of the 
cylinder.  The matrix elements are calculated in Sec.\ \ref{noshear}.  
We will see in Sec.\ \ref{shearres} that the shear flow has the effect 
of mixing nonaxisymmetric disturbances with the axisymmetric varicose 
mode, leading to stabilization in some circumstances.  I will first 
discuss the results for the special case in which the viscosities of 
the two phases are equal, and then generalize to the case of arbitrary 
viscosity ratio.  Some discussion of the relations of this work to 
experiment will be presented in Sections \ref{relexp} and \ref{disc}.

\section{Model equations}

\label{eqtns}

I use the same equations of motion as in \cite{frisch1}.  A simple 
binary mixture can be described by one scalar order parameter $\Phi$, 
the difference in concentration between the two components.  Since we 
are interested in late times after a temperature quench when the 
system consists of well-defined domains, the usual Ginzburg-Landau 
form for the coarse-grained free energy of a symmetrical mixture is 
sufficient to describe the thermodynamics:
	\begin{equation}
	   F[\Phi]=\int d{\bf r} \left(\frac{1}{2} K \left({\bf 
	   \nabla}\Phi\right)^2 - \frac{1}{2}r_0\Phi^2 + 
	   \frac{1}{4}g\Phi^4 \right),
	\end{equation}
where $r_{0}$ and $g$ are positive constants so that the fluid is in 
the two-phase region.  Minimizing the homogeneous part of $F$ leads to 
the values of the concentration in the two bulk phases at equilibrium:
\begin{displaymath}
	\Phi = \pm \sqrt \frac{r_{0}}{g} \equiv \pm \phi_{e}.
\end{displaymath}
The fluid is assumed to be incompressible and sufficiently viscous 
that inertial effects are negligible. The equations of motion for the 
system are then the modified Cahn-Hilliard equation for $\Phi$, the 
Stokes (creeping flow) equation for the velocity field {\bf u}, and 
the incompressibility condition:
	\begin{eqnarray}
		\frac{\partial\Phi}{\partial t} & = & - {\bf u} \cdot {\bf \nabla}\Phi + 
          M\nabla^2 \frac{\delta F}{\delta\Phi},
         \label{CH} \\
         0 & = & \nu  \nabla^{2} {\bf u} + 
			{\bf \nabla} \Phi \frac{\delta F}{\delta\Phi} - {\bf \nabla} P,
		\label{N-S} \\
		0 & = & {\bf \nabla} \cdot {\bf u}.
			\label{incomp1}
  \end{eqnarray}
Here $M$ is a concentration-independent mobility; $\nu$ is the 
viscosity; and $P$ is the pressure, which in general is determined by 
the incompressibility condition (\ref{incomp1}).  The equation for the 
velocity (\ref{N-S}) is generalized to include the coupling of the 
order parameter to the velocity field \cite{chaikin}.  This term leads 
to a capillary force at interfaces, where gradients in $\Phi$ induce 
fluid flow.  Equations (\ref{CH})--(\ref{incomp1}) are the same as 
those of ``model H'' (without the thermal noise terms) used to study 
critical binary fluids \cite{hh}.  These equations have been used 
extensively to study phase separation in binary fluids \cite{bray}.

\begin{figure}
\begin{center}
\leavevmode
\epsfbox{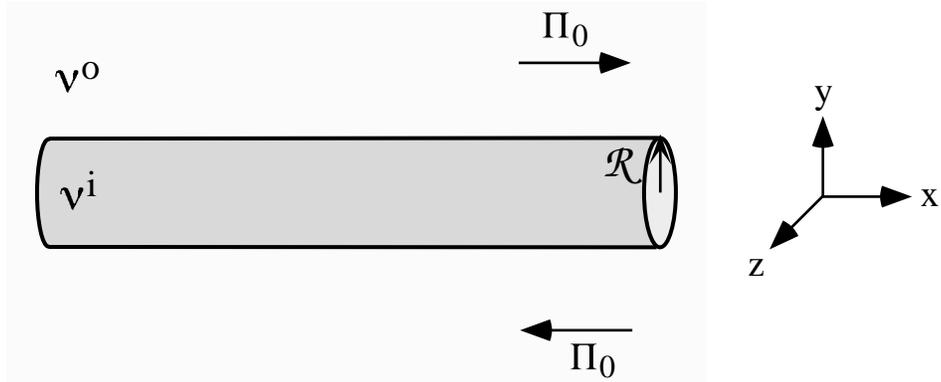}
\end{center}
\caption{\label{cyl} Cylindrical domain in shear flow.}
\end{figure}

Now consider a single cylindrical domain of radius $\cal R$ composed 
of say, phase $\alpha$ with viscosity $\nu^{i}$, immersed in an 
infinite region of phase $\beta$ with viscosity $\nu^{o}$ as 
illustrated in Fig.\ \ref{cyl}.  The external shear flow is imposed 
along the $x$ direction by applying a constant shear stress $\Pi_{0}$ 
far from the cylinder.  Below I will allow for a finite width 
interface between the two phases only in the case that the viscosities 
are equal, $\nu^{i}=\nu^{o}=\nu$; when the viscosities between the two 
phases are different I will assume the interfaces are sufficiently 
sharp that the viscosity changes discontinuously at the interface so 
that Eq.\ (\ref{N-S}) holds in the two different phases separately.  
The first step in a stability analysis of the cylinder is to derive 
the steady-state solutions to the equations of motion which correspond 
to these assumptions and to the geometry of Fig.\ \ref{cyl}.  We 
therefore assume that $\Phi$ is a function of $r$ only and that the 
velocity is only nonzero in the $x$ direction, ${\bf u} = u(r,\theta) 
{\bf \hat{x}}$, and look for time-independent solutions.  The 
Cahn-Hilliard equation (\ref{CH}) has steady state solutions 
satisfying
	   \begin{equation}
			\frac{\delta F}{\delta\Phi} = -K \nabla^{2}\Phi - r_{0}\Phi +
			g \Phi^{3} = \mu = \text{const},
			\label{stat sol}
		\end{equation}
where $\mu$ is the exchange chemical potential.  Using cylindrical 
coordinates $(r,\theta,x)$ the stationary concentration profile 
$\phi_{s}(r)$ therefore satisfies
\begin{equation}
	-K \frac{d^{2}\phi_{s}}{dr^{2}} - 
	\frac{K}{r}\frac{d\phi_{s}}{dr} - r_{0}\phi_{s} + g \phi_{s}^{3} = 
	\mu.
	\label{profile}
\end{equation}
For a sufficiently large radius, $\phi_{s}$ will approach the profile 
for a flat interface between the two coexisting phases,
\begin{equation}
	\phi_{s}(r)  \cong  \phi_{e} \tanh \left[(r-{\cal R})/\xi \right],
	 \quad  \text{large ${\cal R}$},
	\label{profileR} 
\end{equation}
where the width of the interface between the two coexisting phases is 
the thermal correlation length $\xi = \sqrt{2K/r_{0}}$.  I will assume 
throughout that ${\cal R} >> \xi$, so that Eq.\ (\ref{profileR}) is 
reasonable.  Often it will be justified to further approximate the 
interfacial profile by a step function so the interfaces are sharp,
\begin{equation}
	\phi_{s}(r) \approx \phi_{e} \Theta\left[(r-{\cal R})/\xi \right].
	\label{step}
\end{equation}
Note that for either interfacial profile there is a surface tension 
associated with the presence of the interface, which is just the excess 
free energy per unit area at the interface \cite{langer92}:
\begin{equation}
	\sigma = K \int_{-\infty}^{\infty}dr 
	\left(\frac{d\phi_{s}}{dr}\right)^{2} = 
       \frac{4}{3} K \phi_{e}^{2}/\xi = \frac{2}{3} r_{0} \phi_{e}^{2} 
       \xi.
	\label{tension}
\end{equation}

If the viscosities of the two phases are equal, 
$\nu^{i}=\nu^{o}=\nu$, applying a constant shear stress $\Pi_{0}$ 
far from the cylinder leads to simple shear flow everywhere, with 
stationary velocity field
\begin{equation}
	{\bf u}_{s} = \dot{\gamma}y {\bf \hat{x}} = \dot{\gamma} r \cos 
	\theta {\bf \hat{x}},
\end{equation}
where $\dot{\gamma} \equiv \Pi_{0}/\nu$ is the shear rate.  More 
generally, for arbitrary viscosity ratio $\mu \equiv \nu^{i}/\nu^{o}$ 
the stationary velocity field ${\bf u}_{s} = u_{s}(r,\theta) {\bf 
\hat{x}}$ will have a different slope in the two phases.  Taking the 
interface to be mathematically sharp as in Eq.\ (\ref{step}), we can 
solve Equations (\ref{N-S}) and (\ref{incomp1}) for ${\bf u}_{s}$ inside and 
outside the cylinder separately and match the solutions at the 
interface at $r={\cal R}$.  The velocity field must be regular at the 
origin and correspond to simple shear flow far from the cylinder, so 
that
\begin{displaymath}
	\lim_{r\to \infty} {\bf u}_{s} = \dot{\gamma} r \cos \theta {\bf 
	\hat{x}},
\end{displaymath}
where the shear rate is defined in terms of the outer viscosity, 
$\dot{\gamma} \equiv \Pi_{0}/\nu^{o}$. Solving for ${\bf u}_{s}$ gives
\begin{equation}
	{\bf u}_{s}({\bf r}) = \left\{ \begin{array}{ll}
	 \frac{\textstyle 2\dot{\gamma}}{\textstyle \mu + 1} r \cos 
	 \theta {\bf \hat{x}}, & r < {\cal R}  \\
	 \left[ \dot{\gamma}  r  + \left(\frac{\textstyle 2\dot{\gamma} 
	 R^{2}}{\textstyle \mu+1} - \dot{\gamma}R^{2} \right) 
	 \frac{\textstyle 1}{\textstyle r} \right] \cos \theta 
	 {\bf \hat{x}},  & r > {\cal R}.
	 \end{array} \right.
	\label{us}
\end{equation}

It is convenient to rewrite the equations in dimensionless form by 
scaling lengths by the correlation length $\xi$, the concentration by 
its equilibrium magnitude in the bulk phases $\phi_{e}$, and time by 
the characteristic diffusion time $\tau_{\xi}$.  The velocity is 
scaled by the correlation length over the diffusion time:
  \begin{eqnarray*}
		\bar{{\bf r}} & = & {\bf r} \sqrt{\frac{r_0}{2K}}  =  \frac{{\bf 
		r}}{\xi},  \\
		\bar{t} & = &  t \frac{2Mr_{o}}{\xi^{2}} = \frac{t}{\tau_{\xi}}, \\       
		\bar{\Phi} & =  & \frac{\Phi}{\phi_e}, \\
		\bar{{\bf u}} & = &  {\bf u} \frac{\xi}{2Mr_{o}} = {\bf u} 
		\frac{\tau_{\xi}}{\xi}, \\ 
		\bar{P} & = & P \frac{\xi^{2}}{2K\phi_{e}^{2}}. 
	\end{eqnarray*}
Note that the new dimensionless correlation length is $\bar{\xi}=1$.  In 
dimensionless form the equations of motion are now
\begin{equation}
	\frac{\partial\bar{\Phi}}{\partial \bar{t}}  =  - 
	 \bar{{\bf u}} \cdot \bar{{\bf \nabla}}\bar{\Phi} + 
	\frac{1}{2}\bar{\nabla}^2\left(-\frac{1}{2}\bar{\nabla}^{2}\bar{\Phi}
	 - \bar{\Phi}+ \bar{\Phi}^{3}\right),  
\end{equation}
\begin{equation}
	0  = \bar{\nabla}^{2}\bar{{\bf u}} + 
	\frac{1}{\eta} \bar{{\bf \nabla}} \bar{\Phi} \left( 
		-\frac{1}{2}\bar{\nabla}^{2}\bar{\Phi}  - \bar{\Phi}+ \bar{\Phi}^{3}
		 \right) - \frac{1}{\eta} \bar{{\bf \nabla}}\bar{P}, 
\end{equation}
\begin{equation}
    0  =  \bar{{\bf \nabla}} \cdot \bar{{\bf u}}.  
\end{equation}
The equations are characterized by a dimensionless parameter, the 
rescaled viscosity $\eta$:
		\begin{equation}
			\eta = \frac{Mg\nu}{K} = 
			\frac{4Mr_{o}\nu}{3\sigma \xi}. 
		\end{equation}
(In the case of two different viscosites there are two dimensionless 
parameters, $\eta^{i}$ and $\eta^{o}$.)  In dimensionless form the 
stationary solutions corresponding to the cylindrical domain in shear 
are
\begin{equation}
	\bar{\phi}_{s}(\bar{r}) = \tanh (\bar{r}-R),
\end{equation}
\begin{equation}
	\bar{{\bf u}}_{s}(\bar{{\bf r}}) = \left\{ \begin{array}{ll}
	 \frac{\textstyle 2S}{\textstyle \mu + 1} \bar{r} 
	 \cos \theta {\bf \hat{x}}, & \bar{r} < R  \\
	 \left[S  \bar{r}  + \left(\frac{\textstyle 2S
	 R^{2}}{\textstyle \mu+1} - SR^{2} \right) \frac{\textstyle 
	 1}{\textstyle \bar{r}} \right] \cos \theta {\bf \hat{x}}, 
	  & \bar{r} > R, \end{array} \right.
	\label{us2}
\end{equation}
where the dimensionless radius of the cylinder is $R={\cal R}/\xi$.  
The dimensionless shear rate $S \equiv \dot{\gamma} \tau_{\xi}$ is 
simply the product of the shear rate and the diffusion time 
$\tau_{\xi}$ and thus represents a second dimensionless parameter that 
characterizes the strength of the shear flow.

\section{Stability analysis}
\label{stab}

\begin{figure}
\begin{center}
\leavevmode
\epsfbox{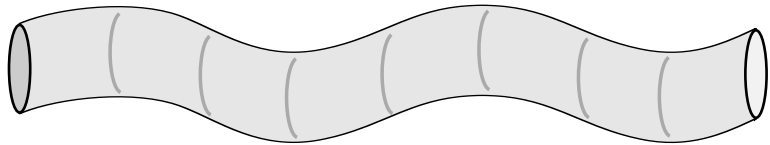}
\end{center}
\caption{\label{m1} Undulation $m=1$ mode of the cylinder.}
\end{figure}

In this section I present the strategy for calculating the stability 
of the cylindrical domain.  We know that an infinite cylinder is 
unstable to varicose perturbations.  One could imagine other, 
nonaxisymmetric perturbations of the cylinder as well, such as the 
``undulation'' mode shown in Fig.\ \ref{m1}.  In Sec.\ \ref{noshear} 
we will find that the cylinder is stable to all of these 
perturbations.  The shear flow will have the effect of mixing the 
different possible perturbations.

Consider small perturbations about the stationary solutions found 
above (in the rest of the discussion I will drop the bars over the 
dimensionless variables for clarity):
\begin{eqnarray}
			\phi & = & \Phi - \phi_s,  \\
			{\bf v} & = & {\bf u - u}_s.
		\end{eqnarray}
To linear order in the small perturbations $\phi$ and ${\bf v}$, the 
equations of motion are
\begin{eqnarray}
	\frac{\partial \phi}{\partial t} & = & -u_{s}(r) \cos \theta 
	\frac{\partial \phi}{\partial x} - \phi_{s}'(r)v_{r} \nonumber \\
	& &  + \frac{1}{2} \nabla^{2} \left(-\frac{1}{2}\nabla^{2} + 
	W_{s}(r)\right) \phi,
	\label{conc}
\end{eqnarray}
\begin{equation}
	0 = \nabla^2{\bf v} + \frac{1}{\eta} \phi_{s}'(r)
         	\left(-\frac{1}{2}\nabla^2+W_s(r) \right) \phi 
         	{\bf \hat{r}} - \frac{1}{\eta} {\bf \nabla} P,   
	\label{hydro}
\end{equation}
\begin{equation}
	0 = {\bf \nabla} \cdot {\bf v}.
	\label{incomp}
\end{equation}
Here $\eta$ is the appropriate viscosity for whichever region is under 
consideration, $v_{r}$ is the $r$ component of the perturbed velocity 
field {\bf v}, and primes indicate differentiation with respect to 
$r$. $W_{s}$ is a function of the stationary concentration profile:
 \begin{equation}
	      W_s(r) = \left. \frac{\partial^2 f}{\partial \phi^2} 
	      \right|_{{\displaystyle \phi_s(r)}} = -1 + 3 \phi_{s}^{2}(r).
  \end{equation}

The time dependence of the perturbations is determined by the 
concentration equation (\ref{conc}).  The system is translationally 
invariant in the $x$ direction, so we can write any perturbation as a 
sum over Fourier modes in $x$.  Since we are interested in the growth (or 
damping) of perturbations we take
\begin{equation}
	\phi = \phi(r,\theta) e^{ikx-\omega t}, \quad {\bf 
	v} = {\bf v}(r,\theta) e^{ikx-\omega t}.
	\label{fourier}
\end{equation}
We will be interested in long-wavelength fluctuations for which 
the dimensionless wave number $k <<1$ (let $\kappa=k/\xi$ be the 
wave number and $\varpi = \omega/\tau_{\xi}$ be the 
growth (damping) rate in the original variables).  Substitution 
into Eq.\ (\ref{conc}) leads to an eigenvalue equation for the 
growth rate $\omega$:
\begin{eqnarray}
	 \omega \phi(r,\theta) & = & iku_{s}(r) \cos \theta \phi(r,\theta) + 
	\phi_{s}'(r) v_{r}(r,\theta)  \nonumber \\
	& & - \frac{1}{2} \left(\frac{1}{r}\frac{\partial}{\partial 
	r}\left(r \frac{\partial}{\partial r}\right) + 
	\frac{1}{r^{2}}\frac{\partial^{2}}{\partial \theta^{2}} - 
	k^{2}\right) \nonumber \\
	& & \times \left(-\frac{1}{2}\frac{1}{r}\frac{\partial}{\partial 
	r}\left(r \frac{\partial}{\partial r}\right) - 
	\frac{1}{2r^{2}}\frac{\partial^{2}}{\partial \theta^{2}} 
	\right. \nonumber \\
	& & \left. + 
	\frac{1}{2} k^{2}+ W_{s}(r)\right) \phi(r,\theta).
	\label{ev}
\end{eqnarray}
A real, positive value of $\omega(k)$ indicates stability of the 
cylinder against the perturbation.  Since $v_{r}$ depends on $\phi$ 
through Eq.\ (\ref{hydro}), this eigenvalue equation is essentially an 
integro-differential equation in which the expression for $v_{r}$ acts 
as an integral operator on $\phi$.

Eq.\ (\ref{ev}) cannot be solved exactly, so we need an approximate 
approach.  Following the calculational approach outlined in 
\cite{frisch1}, first consider the Cahn-Hilliard part of Eq.\ 
(\ref{ev}), without the hydrodynamic terms:
\begin{equation}
	\omega \phi = {\bf \Gamma} {\bf F} \phi,
	\label{CH2}
\end{equation}
where we have defined the operators
\begin{mathletters}
\begin{equation}
		{\bf \Gamma} = - \frac{1}{2} \left(\frac{1}{r}\frac{\partial}{\partial 
	r}\left(r \frac{\partial}{\partial r}\right) + 
	\frac{1}{r^{2}}\frac{\partial^{2}}{\partial \theta^{2}} - 
	k^{2}\right),
\end{equation}
\begin{equation}
	 {\bf F} =  \left(-\frac{1}{2}\frac{1}{r}\frac{\partial}{\partial 
	r}\left(r \frac{\partial}{\partial r}\right) - 
	\frac{1}{2r^{2}}\frac{\partial^{2}}{\partial \theta^{2}} + \frac{1}{2} 
	k^{2}+ W_{s}(r)\right).
\end{equation}
\end{mathletters}
This part of Eq.\ (\ref{ev}) includes the dynamics of the 
concentration field on the scale of the interface.  Since $S<1$ 
($\dot{\gamma} \tau_{\xi}<1$), the shear flow acts on the scale of the 
domains but is not strong enough to alter the interfacial profile.  As 
we will see shortly, this assumption allows us to find an approximate 
solution for $\phi$.  Let $\phi_{n}$ be the set of eigenfunctions of 
Eq.\ (\ref{CH2}) and define a set of ``conjugate'' functions by
\begin{equation}
	{\bf \Gamma} \tilde{\phi_{n}} = \phi_{n}.
\end{equation}
Then one can show that ${\bf \Gamma}$ and ${\bf F}$ are Hermitian 
operators (although their product is not) as long as the $\phi_{n}$ 
and $\tilde{\phi_{n}}$ obey periodic boundary conditions or vanish at 
infinity.  The eigenvalues $\omega_{n}$ are real and the 
eigenfunctions and their conjugates are orthogonal:
\begin{displaymath}
	(\tilde{\phi}_{m}, \phi_{n}) \equiv \int d{\bf r} 
	\tilde{\phi}_{m}^{*}({\bf r}) \phi_{n}({\bf r}) = 0 \quad \rm{for} 
	\quad n \neq m.
\end{displaymath}
For any pair of trial functions $\phi_{0}$ and $\tilde{\phi}_{0}$ 
obeying the same boundary conditions, there is a variational relation 
which gives an upper bound on the lowest eigenvalue $\omega$
\cite{langer71,jasnow}:
	\begin{equation}
		\omega_{min} \leq \frac{(\phi_0, {\bf F}\phi_0)}{(\tilde{\phi}_0, 
		\phi_0)}.
		\label{var}
	\end{equation}
Here the parentheses again indicate inner products.  

This variational theorem can be exploited to find solutions to Eq.\ 
(\ref{CH2}) corresponding to various perturbations of the cylinder.  
Application of Eq.\ (\ref{var}) requires a good trial function 
$\phi_{0}$.  The smallest eigenvalues of Eq.\ (\ref{CH2}) will 
correspond to eigenfunctions describing $\theta$-dependent 
deformations of the cylinder, in which the interface is translated by 
a small amount but the interfacial width remains fixed 
\cite{frisch1,langer71}.  Higher eigenvalues correspond to other 
deformation modes in which the structure of the interface changes, 
such as breathing modes which change the width of the interface.  I 
will neglect all such modes here, since they are more quickly damped 
than the slow translational modes and are not important to the 
dynamics on the scale $R$.  Thus, we can solve the Cahn-Hilliard part 
of the eigenvalue equation (\ref{ev}) by using the variational theorem 
with a trial function corresponding to the translational deformation of 
interest.

The translational modes can be characterized by their angular 
dependences.  Any general perturbation of the concentration field can 
be expanded as a Fourier series in $\theta$:
\begin{equation}
	\phi(r,\theta)  =  \sum_{m} e^{im\theta}\phi_{m}(r),
	\label{phitheta}  \\
\end{equation}
where $\phi_{m}(r)$ is the function necessary to translate the 
interface by a small amount $\delta r$ in the ${\bf \hat{r}}$ 
direction (the functional form of the $\phi_{m}(r)$ will be calculated 
below in Sec.\ \ref{diffusive}).  The Cahn-Hilliard part of the 
eigenvalue equation can then be rewritten as
\begin{equation}
	\omega_{m} \phi_{m}(r) = {\bf \Gamma}_{m}(r) {\bf F}_{m}(r) 
	\phi_{m}(r),
	\label{CHm}
\end{equation}
where the operators are
\begin{mathletters}
\begin{eqnarray}
	{\bf \Gamma}_{m} & = & -\frac{1}{2} \frac{d^{2}}{dr^{2}} 
	-\frac{1}{2r}\frac{d}{dr} + \frac{m^{2}}{2r^{2}} + 
	\frac{1}{2}k^{2}, \\
	{\bf F}_{m} & = & -\frac{1}{2} \frac{d^{2}}{dr^{2}}
	 -\frac{1}{2r}\frac{d}{dr} + \frac{m^{2}}{2r^{2}} + \frac{1}{2}k^{2} + 
	3\phi_{s}^{2} - 1.
	\label{Fm}
\end{eqnarray}
\end{mathletters}
Each mode $\phi_{m}(r)e^{im\theta}$ corresponds to a different 
geometrical perturbation mode of the cylinder.  In the absence of the 
external shear flow, the cylindrical domain will be unstable to the 
axisymmetric, $m=0$ varicose mode as discussed in the introduction 
(see Fig.\ \ref{varicose}).  We will see below that the $m=1$ mode 
shown in Fig.\ \ref{m1} is an exact solution to Eq.\ (\ref{CH2}) for 
$k=0$.  At $k=0$, it simply corresponds to a uniform translation of 
the entire cylinder and is thus marginally stable with eigenvalue 
$\omega_{1}(k=0)=0$.  We might anticipate that the cylinder will be 
stable to higher modes in $m$ as well.  Note that the shear flow term 
in Eq.\ (\ref{ev}) is proportional to $\cos \theta$, so this term 
should have the effect of mixing modes with different values of $m$.

Now consider the full eigenvalue equation, Eq.\ (\ref{ev}).  We are 
interested in the stability of perturbations characterized by the 
various $\theta$-dependent translational modes.  The functions 
$\phi_{m}(r)e^{im\theta+ikx}$ are approximate eigenvectors of Eq.\ 
(\ref{CH2}).  To solve the full equation we adopt an approximation 
similar to ``tight-binding'' or $k \cdot p$ approximations used in 
solid state physics.  We assume the translational modes are good basis 
states for the full problem and write Eq.\ (\ref{ev}) as a matrix 
equation in this basis.  We can truncate the matrix to only include a 
finite number of states $m$ and then diagonalize the matrix to find 
the stability eigenvalues.  This is valid when the two hydrodynamic 
terms are small enough that they only cause mixing among the $m$ 
states included in the basis; they must be small relative to the 
distance to the next higher eigenvalue not included.  Also, the shear 
flow must satisfy $S<1$ so that it is reasonable to only consider the 
translational deformation modes.  In the strong-shear regime $S>1$ 
($\dot{\gamma} \tau_{\xi}>1$) the shear flow might couple to other 
modes that we have neglected, which alter the width or shape of the 
interfacial profile itself, since these modes only damp out on a time 
scale of roughly $\tau_{\xi}$.  Note that the term containing 
$v_{r}(r,\theta)$ in Eq.\ (\ref{ev}) depends on $\phi$ through the 
hydrodynamic equation (\ref{hydro}).  So for each mode $m$ we can 
solve Eq.\ (\ref{hydro}) for $v_{r}(r,\theta)$, assuming that 
$\phi(r,\theta)$ is given by the approximate basis function 
$\phi_{m}(r) e^{im\theta}$.  In general the resulting velocity field 
can then also be expanded as a Fourier series.  Denoting by ${\bf 
v}_{m}$ the solution for ${\bf v}$ obtained from substituting Eq.\ 
(\ref{fourier}) and $\phi(r,\theta)=\phi_{m}(r) e^{im\theta}$ into 
Eq.\ (\ref{hydro}), we can expand
\begin{equation}
	{\bf v}_{m}(r,\theta) = \sum_{n} e^{in\theta}{\bf v}_{nm}(r).
	\label{vnm}
\end{equation}
The $\theta$ dependence of ${\bf v}$ will not necessarily be the 
same as that of $\phi_{m}$ so in general the coefficients ${\bf 
v}_{nm}(r)$ in the sum will be nonzero even for $n \neq m$.

To obtain the effective matrix equation corresponding to Eq.\ 
(\ref{ev}) we write $\phi(r,\theta)$ as a vector
\begin{displaymath}
	\phi(r,\theta) = \left( \begin{array}{c}
	    \epsilon_{0} \phi_{0}(r) \\
	    \epsilon_{1} \phi_{1}(r) e^{i\theta} \\
	    \epsilon_{2} \phi_{2}(r) e^{2i\theta} \\
	    \vdots   
	   \end{array} \right) 
\end{displaymath}
and multiply on the left in Eq.\ (\ref{ev}) by the corresponding 
conjugate vector.  Here the $\epsilon_{m}$ are the amplitudes of the 
small perturbations $\phi_{m}$.  Recall the conjugate function 
$\tilde{\phi}$ is defined by ${\bf \Gamma} \tilde{\phi} = \phi$ so it 
satisfies the Poisson equation
\begin{equation}
	-\frac{1}{2} \nabla^{2} \tilde{\phi} = \phi.
\end{equation}
We can expand $\tilde{\phi}$ in the same way as $\phi$ so that
\begin{displaymath}
	\tilde{\phi}(r,\theta,x,t) = \sum_{m} \tilde{\phi}_{m}(r) e^{im\theta + ikx} 
	e^{-\omega t},
\end{displaymath}
in which case ${\bf \Gamma}_{m} \tilde{\phi}_{m} = \phi_{m}$.  We can 
easily solve for $\tilde{\phi}_{m}(r)$ using the Green's function for 
the Laplacian in cylindrical coordinates.  The result is
\begin{equation}
	\tilde{\phi}_{m}(r) = \int_{0}^{\infty} r' dr' 2 K_{m}(kr_{>}) 
	I_{m}(kr_{<}) \phi_{m}(r'),
\end{equation}
where $r_{<}$ ($r_{>}$) indicates the lesser (greater) of $r$ and 
$r'$, and $K_{m}, I_{m}$ are the modified Bessel functions.  Then Eq.\ 
(\ref{ev}) becomes after multiplying on the left by
\begin{displaymath}
	\tilde{\phi}(r,\theta)^{*} = (\begin{array}{cccc} 
	\tilde{\phi}_{0}(r) & \tilde{\phi}_{1}(r) e^{-i\theta} & 
	\tilde{\phi}_{2}(r) e^{-2i\theta} & \cdots \end{array})
\end{displaymath}
and integrating over all $\theta$,

\end{multicols}
\setlength{\unitlength}{1in}
\begin{picture}(3.5,.25)
\put(0,.125){\line(1,0){3}}
\end{picture}

\begin{eqnarray}
	\lefteqn{ \left(\begin{array}{cccc}
	(\tilde{\phi_{0}}, \phi_{0}) \omega & 0 & 0 & \\
	0 & (\tilde{\phi_{1}}, \phi_{1})\omega & 0 & \cdots \\
    0 & 0 & (\tilde{\phi_{2}}, \phi_{2})\omega & \\
	 & \vdots &  & \\
	\end{array} \right)
	\left(\begin{array}{c} \epsilon_{0} \\ \epsilon_{1} \\ \epsilon_{2} 
	\\ \vdots  \end{array} \right)  =  } \nonumber \\ 
	 & & \left(\begin{array}{cccc}
	(\tilde{\phi}_{0},v_{r,00}\phi_{s}') + 
	(\phi_{0},{\bf F}_{0}\phi_{0}) & 
	(\tilde{\phi}_{0},\frac{1}{2}iku_{s} \phi_{1} + 
	v_{r,01}\phi_{s}') & 0 & \\
	(\tilde{\phi}_{1}, \frac{1}{2}iku_{s} \phi_{0} + v_{r,10}\phi_{s}')  &
	(\tilde{\phi}_{1},v_{r,11}\phi_{s}') +(\phi_{1}, {\bf 
	F}_{1}\phi_{1}) & (\tilde{\phi}_{1}, \frac{1}{2}iku_{s} 
	\phi_{2} + v_{r,12}\phi_{s}')  & \cdots \\
	0 & (\tilde{\phi}_{2}, \frac{1}{2}iku_{s} \phi_{1} +
	v_{r,21}\phi_{s}')  & 
	(\tilde{\phi}_{2},v_{r,22}\phi_{s}') +(\phi_{2}, {\bf 
	F}_{2}\phi_{2}) & \\
	 & \vdots & & \\
	\end{array} \right) 
		\left(\begin{array}{c} \epsilon_{0} \\ \epsilon_{1} \\ \epsilon_{2} 
	\\ \vdots  \end{array} \right). 
	\label{matrix1}
\end{eqnarray}

\begin{picture}(6.5,.25)
\put(3.75,.125){\line(1,0){3}}
\end{picture}

\begin{multicols}{2}
Solving this equation gives approximate stability eigenvalues $\omega(k)$ for 
the cylinder in the shear flow.  We have used the definition ${\bf 
\Gamma}_{m}\tilde{\phi}_{m} = \phi_{m}$; the diagonal diffusive 
terms are then exactly the variational expression (\ref{var}).  
The shear terms involving the stationary velocity $u_{s} \propto 
S$ are completely off-diagonal, so they will indeed have the 
effect of mixing the modes.  We will find in Sec.\ \ref{vmu} that 
the off-diagonal elements involving ${\bf v}$ only mix modes that 
differ by $\pm 1$ as written in Eq.\ (\ref{matrix1}) and that they 
are also directly proportional to the shear rate $S$.  Thus in the 
absence of the shear flow, $S=0$, the matrix is completely 
diagonal and the $m$ modes are independent, with stability 
eigenvalues
\begin{eqnarray}
	\omega_{m} & = & \frac{(\tilde{\phi}_{m}, v_{r,m}\phi_{s}') + 
	(\phi_{m}, {\bf F}_{m} \phi_{m})}{(\tilde{\phi}_{m},\phi_{m})} 
	\nonumber \\ 
	& \equiv & \omega_{m,h} + \omega_{m,d}, \quad S=0.  
	\label{evm}
\end{eqnarray}
These zero-shear stability eigenvalues are the sum of two terms, one 
due to hydrodynamic transport in the system and the other due to 
diffusive transport.  Solving Eq.\ (\ref{matrix1}) for non-zero shear 
rate requires truncating the matrix at some mode $m$; since we expect 
only the $m=0$ mode to be (possibly) unstable, we might anticipate 
that only a few of the higher modes are needed to investigate the 
behavior of the $m=0$ mode under shear.

To summarize the results of this section, the equations of motion were 
first linearized in the small perturbations $\phi$ and ${\bf v}$ and 
expressed parametrically in terms of the wave number $k$.  The 
perturbations of the cylinder of interest here, the translational 
modes, were characterized by their dependence on $\theta$.  A 
variational expression was introduced for the diffusive part of the 
problem, and the eigenvalue equation for the full problem was written 
as a matrix equation in the basis of the $m$ translation modes.  In 
the remaining sections I calculate the various matrix elements in Eq.\ 
(\ref{matrix1}), which requires solving the hydrodynamic equation for 
the perturbed velocity field ${\bf v}$, and then solve the matrix 
equation itself and examine the results for various parameters of 
interest.

\section{Matrix elements and results without shear}
\label{noshear}

\subsection{Diffusive contribution}
\label{diffusive}

We begin by calculating the diffusive contribution to the matrix 
elements $\omega_{m,d}$ as defined in Eq.\ (\ref{evm}).  We need 
to determine the $r$ dependence of the basis functions $\phi_{m}$.  
For $v_{r}=0$ and $k=0$, translating the entire cylindrical 
interface by an amount $d{\bf r}$, $\phi_{s}({\bf r}+d{\bf r}) = 
\phi_{s}(r)+{\bf \nabla} \phi_{s} \cdot d{\bf r}$, requires adding the 
function
\begin{equation}
	 {\bf \nabla} \phi_{s} = \phi_{s}' {\bf \hat{r}} = \phi_{s}'(r)(\cos \theta 
	{\bf \hat{y}} + \sin \theta {\bf \hat{z}}) 
\end{equation}
to the original interfacial profile.  But the translated interface, 
and therefore ${\bf \nabla} \phi_{s}$, should also be an exact solution of 
(\ref{CHm}).  We can verify this by differentiating Eq.\ (\ref{profile}) 
for the stationary solution $\phi_{s}$ with respect to $r$:
\begin{equation}
	-\frac{1}{2} \frac{d^{3}\phi_{s}}{dr^{3}} - 
	\frac{1}{2r}\frac{d^{2}\phi_{s}}{dr^{2}} + \frac{1}{2r^{2}} 
	\frac{d\phi_{s}}{dr} - \frac{d\phi_{s}}{dr} + 3\phi_{s}^{2} 
	\frac{d\phi_{s}}{dr} = 0.
\end{equation}
But this is simply equivalent to 
\begin{equation}
	{\bf F}_{m=1} \phi_{s}' = 0,
\end{equation}
so $\phi_{1}(r) \equiv \phi_{s}'(r)$ is an exact eigenfunction of 
${\bf \Gamma}_{1}{\bf F}_{1}$ for $k=0$, with eigenvalue 
$\omega_{1}=0$.  We can exploit this solution to approximate ${\bf 
F}_{m}\phi_{m}$ for general $m$.  Since $R \gg 1$ ($\cal{R} \gg \xi$) 
and since $\nabla^{2} \phi$ is only significant near the interface at 
$r=R$, we can in general replace the term $m^{2}/2r^{2}$ by 
$m^{2}/2R^{2}$ in the expression for ${\bf F}_{m}$ (\ref{Fm}).  But then 
we would expect all $m$ modes to have roughly the same radial 
dependence as the $m=1$ mode, so we can approximate
\begin{equation}
	{\bf F}_{m}\phi_{m} \approx {\bf F}_{m}\phi_{s}' =
	{\bf F}_{m=1}\phi_{s}' +  \left(\frac{m^{2}}{2r^{2}} - 
	\frac{1}{2r^{2}} \right) \phi_{s}'.
\end{equation}
For nonzero $k$, ${\bf F}_{m=1}\phi_{s}' = \frac{1}{2}k^{2}\phi_{s}'$, 
so this gives
\begin{equation}
	{\bf F}_{m}\phi_{m} \approx \left(\frac{1}{2} k^{2} + 
	\frac{m^{2}-1}{2r^{2}}\right) \phi_{s}'.
	\label{fmphim}
\end{equation}
From Eq.\ (\ref{evm}) the diffusive part of the diagonal matrix 
element $\omega_{m}$ is then
\begin{equation}
	\omega_{m,d} \cong \frac{1}{(\tilde{\phi}_{m},\phi_{m})} 
	\int_{0}^{\infty} r\, dr\, \phi_{m} \left(\frac{1}{2} k^{2} + 
	\frac{m^{2}-1}{2r^{2}}\right)  \phi_{s}'.
\end{equation}
We are assuming we can approximate $\phi_{s}'$ by the flat interface 
form Eq.\ (\ref{profileR}) so $\phi_{s}' \cong {\mathrm sech}^{2}(r-R)$.  
Since all modes have the same radial dependence we set $\phi_{m}(r) = 
\phi_{s}'(r)$ in the denominator as well, which gives
\begin{eqnarray}
	\omega_{m,d} & \cong &
	\frac{1}{(\tilde{\phi}_{m}(r),\mathrm{sech}^{2}(r-R))} 
	\int_{0}^{\infty} dr \left(\frac{1}{2} k^{2}r \right. \nonumber \\
	& & \left. + \frac{m^{2}-1}{2r}\right) \mathrm{sech}^{4}(r-R),
	\label{wmdf}
\end{eqnarray}
\begin{equation}
	\tilde{\phi}_{m}(r) = \int_{0}^{\infty} r'\, dr'\, 2 K_{m}(kr_{>}) 
	I_{m}(kr_{<}) \mathrm{sech}^{2}(r'-R).
	\label{tphimr}
\end{equation}
The diffusive contribution to $\omega$ has thus been reduced to 
quadrature.

\begin{figure}
\begin{center}
\leavevmode
\epsfxsize= 4in
\epsfbox{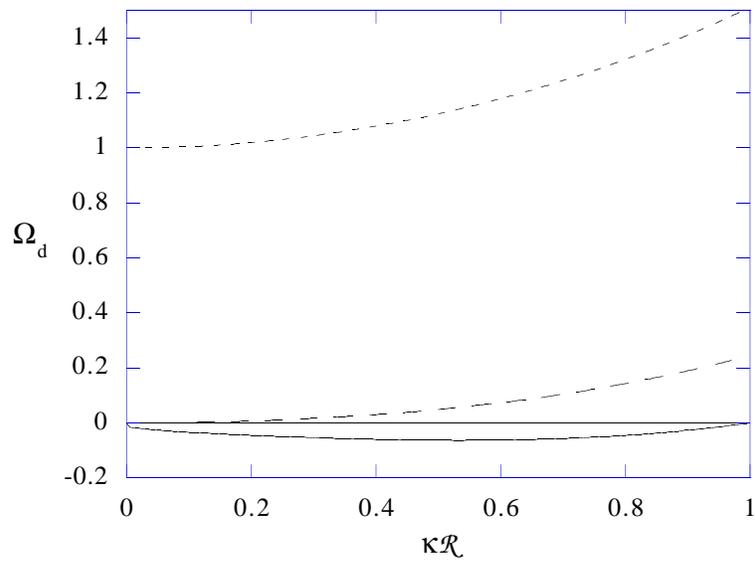}
\end{center}
\caption{\label{wdg0} Dispersion relations in no shear, 
$\Omega_{0,d}$ (solid curve), $\Omega_{1,d}$ (dashed curve), and 
$\Omega_{2,d}$ (dotted curve).}
\end{figure}

Equations (\ref{wmdf}) and (\ref{tphimr}) are integrable numerically and 
are valid for diffuse interfaces of width $\xi$.  However, 
qualitatively the results are the same for sharp interfaces, in which 
case the result can be expressed in closed form.  The easiest way to 
take the sharp interface limit is to take
\begin{eqnarray}
	{\mathrm sech}^{2}(r-R) & \rightarrow & 2 \delta(r-R), \nonumber \\
	{\mathrm sech}^{4}(r-R) & \rightarrow & \frac{4}{3}\delta(r-R),
	\label{sharpint}
\end{eqnarray}
in all of the integrals, where $\delta(r-R)$ is the Dirac delta 
function.  (This is equivalent to using the step-function profile Eq.\ 
(\ref{step}).)  The prefactors maintain the correct normalizations, so 
we are taking the width of the interface $\xi \rightarrow 0$ while 
keeping the surface tension $\sigma$ constant.  The delta functions 
then allow easy evaluation of the necessary integrals.  The conjugate 
function $\tilde{\phi}_{m}$ is
\begin{equation}
	\tilde{\phi}_{m}(r) \cong  \left\{ \begin{array}{cc}
	    4RK_{m}(kR)I_{m}(kr), & r<R \\
	    4RK_{m}(kr)I_{m}(kR), & r>R, \end{array} \right. 
\end{equation}
so the normalization integral is
\begin{eqnarray}
	(\tilde{\phi}_{m}, \phi_{m}) & \cong & \int_{0}^{\infty} r\, dr\, 
	\tilde{\phi}_{m}(r) 2\delta (r-R) \nonumber \\
	& \cong & 8 R^{2} K_{m}(kR)I_{m}(kR).
\end{eqnarray}
Substituting into Eq.\ (\ref{wmdf}) we find the diffusive term in 
the stability eigenvalue is
\begin{equation}
	\omega_{m, d} 
	  =  \frac{k^{2}R^{2}+m^{2}-1}{12R^{3}K_{m}(kR)I_{m}(kR)}.
	 \label{wmd}
\end{equation}
In the original variables, the dispersion relation is
\begin{equation}
	\varpi_{m,d} \tau_{\xi} = 
	\frac{\kappa^{2}{\cal{R}}^{2}+m^{2}- 1} 
	{12({\cal{R}}^{3}/\xi^{3}) K_{m}(\kappa{\cal{R}})I_{m}(\kappa{\cal 
	R})},
	\label{wmds}
\end{equation}
which we note can be written independently of $\xi$ as
\begin{equation}
	\varpi_{m,d} = \frac{M\sigma}{4\phi_{e}^{2}} 
	\frac{\kappa^{2}{\cal{R}}^{2}+m^{2}- 1} 
	{{\cal{R}}^{3} K_{m}(\kappa{\cal{R}})I_{m}(\kappa {\cal R})}.
\end{equation}
We immediately see that the axisymmetric $m=0$ mode is 
thermodynamically unstable for $\kappa{\cal R} <1$, {\em i.e.}\ for 
wavelengths longer than the circumference of the cylinder, whereas 
the $m=1$ translation mode is stable for all $\kappa$ as 
predicted.  Fig.\ \ref{wdg0} shows the dimensionless part of the 
first three modes,
\begin{equation}
	\Omega_{m,d}(\kappa {\cal R}) \equiv 
	\frac{\kappa^{2}{\cal{R}}^{2}+m^{2}- 1} {12
	K_{m}(\kappa{\cal{R}})I_{m}(\kappa {\cal R})},
	\label{Omd}
\end{equation}
where $\varpi_{m,d}\tau_{\xi} = \xi^{3} \Omega_{m,d}(\kappa {\cal 
R})/{\cal R}^{3}$.  These are the diffusive contributions to the 
stability eigenvalues in the absence of the shear flow.

\begin{figure}
\begin{center}
\leavevmode
\epsfxsize= 4in
\epsfbox{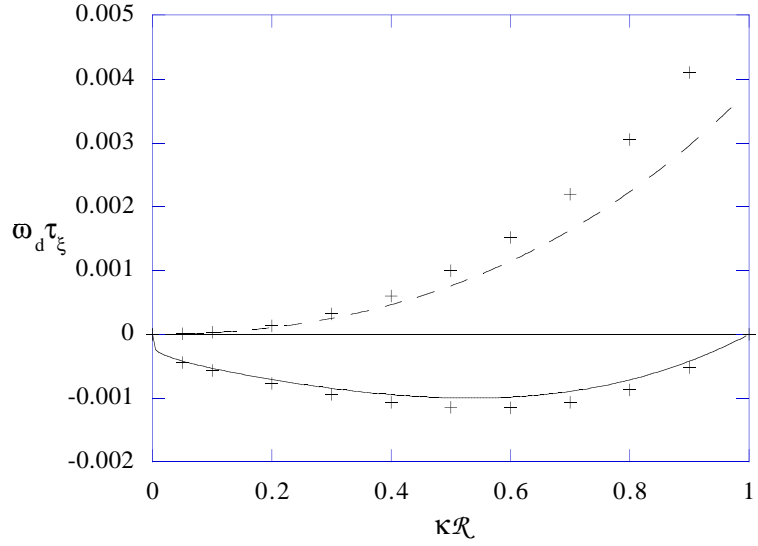}
\end{center}
\caption{\label{finite} Corrections to the eigenvalues for sharp 
interfaces $\varpi_{0,d}$ (solid curve) and $\varpi_{1,d}$ (dashed 
curve) due to a finite width interface ($+$ symbols), for ${\cal 
R}=4\xi$.}
\end{figure}

Fig.\ \ref{finite} shows the difference between using finite width 
interfaces in Eq.\ (\ref{wmdf}) and using the sharp interface 
expression (\ref{wmds}) for a cylinder of radius ${\cal R}=4\xi$, for 
the lowest two eigenvalues, $\varpi_{0,d}$ and $\varpi_{1,d}$.  
The finite width curve is at most $15\%$ more negative than the
sharp interface curve for $\varpi_{0,d}$, and $31\%$ larger than 
the sharp interface curve for $\varpi_{1,d}$, at $\kappa {\cal R} = 
.5$ (near the maximally unstable varicose mode).  The shapes of the 
curves are qualitatively the same in both cases.  As $\cal{R}$ 
increases and the step-function approximation for the interfacial 
profile becomes better, the difference between the curves grows 
smaller.  For ${\cal R}=6\xi$ the differences are reduced to roughly 
$9\%$ and $20\%$, respectively.  Thus the qualitative behavior should 
remain unchanged for a domain with sharp interfaces, but detailed
comparison with experimental or simulational data may require 
including the effect of having diffuse interfaces.

\subsection{Velocity field, equal viscosities}

Next consider the matrix elements involving the perturbed velocity 
field ${\bf v}$.  This section will be limited to the special case in 
which the viscosities of the two phases are equal, $\eta^{i} = 
\eta^{o} = \eta$.  Then it will turn out that the velocity matrix 
elements are completely diagonal, even with the shear flow, and it 
will be possible to obtain closed form expressions for ${\bf 
v}_{mm}(r) \equiv {\bf v}_{m}(r)$.  

Recall that the perturbed velocity field ${\bf v}$ satisfies Equations 
(\ref{hydro}) and (\ref{incomp}):
\begin{equation}
	0 = \nabla^{2} {\bf v} - \frac{1}{\eta} {\bf \nabla}P + 
	\frac{1}{\eta} \phi_{s}' {\bf F} \phi {\bf \hat{r}},
	\label{hydro2}
\end{equation}
\begin{equation}
	0 = {\bf \nabla} \cdot {\bf v},
	\label{incomp2}
\end{equation}
where ${\bf F} = -\frac{1}{2}\nabla^{2} + W_{s}$.  To solve for ${\bf 
v}$, we follow a general procedure from Happel and Brenner 
\cite{happel}.  Taking the divergence of Eq.\ (\ref{hydro2}) and 
applying the incompressibility condition (\ref{incomp2}) leads to a 
Poisson equation for the pressure $P$:
\begin{equation}
	\nabla^{2} P = {\bf \nabla} \cdot (\phi_{s}' {\bf F} \phi {\bf 
	\hat{r}}).
\end{equation}
Expanding the pressure as $P = P_{m}(r) e^{im\theta + ikx} e^{-\omega t}$
we find $P_{m}(r)$ satisfies
\begin{equation}
	P_{m}'' + \frac{1}{r}P_{m}' - \frac{m^{2}}{r^{2}}P_{m} - k^{2}P_{m}
	 = \frac{1}{r} \frac{d}{dr}(r \phi_{s}' {\bf F}_{m} \phi_{m}),
\end{equation}
the homogeneous part of which is simply the modified Bessel
equation. Using a Green's function and requiring $P_{m}(r)$ to be finite at 
$r=0$ and to vanish as $r \rightarrow \infty$, we have
\begin{eqnarray}
	\lefteqn{P_{m}(r)  = } \nonumber \\
	& & -\int_{0}^{\infty}\!r'\, dr' \, K_{m}(kr_{>})I_{m}(kr_{<}) 
	\frac{1}{r'} \frac{d}{dr'}\left[r' \phi_{s}' {\bf F}_{m} 
	\phi_{m}(r')\right] 
	 \nonumber \\
	&  & \cong \int_{0}^{\infty}dr' \, \frac{d}{dr'}\left[ 
	K_{m}(kr_{>})I_{m}(kr_{<})\right] \left(\frac{1}{2} k^{2}r' \right.
	\nonumber \\ 
	& & \left.  \quad \quad + \frac{m^{2}-1}{2r'} \right) {\mathrm 
	sech}^{4}(r'-R),
	\label{pressure}
\end{eqnarray}
where we have used Eq.\ (\ref{fmphim}).  Substituting this 
expression for $P$ into Eq.\ (\ref{hydro2}) gives a vector Poisson 
equation for {\bf v}.

In cylindrical coordinates, the $r$ and $\theta$ components of ${\bf v}$ are 
coupled:
\begin{mathletters}
\label{vcoupled}
\begin{eqnarray}
	\nabla^{2} v_{r} - \frac{2}{r^{2}}\frac{\partial 
	v_{\theta}}{\partial \theta} - \frac{v_{r}}{r^{2}} & = & \frac{1}{\eta} 
	\frac{\partial P}{\partial r} - \frac{1}{\eta} \phi_{s}' {\bf F} 
	\phi,
	\label{rho}  \\
	\nabla^{2} v_{\theta} + \frac{2}{r^{2}}\frac{\partial 
	v_{r}}{\partial \theta} - \frac{v_{\theta}}{r^{2}} & = & 
	\frac{1}{\eta r} \frac{\partial P}{\partial \theta}.
	\label{Theta}
\end{eqnarray}
\end{mathletters}
We can solve for both together by writing
\begin{mathletters}
\label{vexp}
\begin{eqnarray}
    v_{r,m} & = & \varrho_{m}(r) e^{im\theta} e^{ikx-\omega t}, \\
    v_{\theta,m} & = & -i \vartheta_{m}(r)  e^{im\theta} e^{ikx-\omega 
    t},
\end{eqnarray}
\end{mathletters}
so that $v_{\theta}$ has a phase shift relative to $v_{r}$.  Eqs.\ 
(\ref{vcoupled}) then become
\begin{mathletters}
\begin{eqnarray}
    & \varrho_{m}'' + \frac{1}{r}\varrho_{m}' - \left(k^{2} + 
    \frac{m^{2}+1}{r^{2}} \right) \varrho_{m} - \frac{2m}{r^{2}} 
    \vartheta_{m}  = & \nonumber \\
    &  \frac{1}{\eta} P_{m}' - 
    \frac{1}{\eta} \phi_{s}' {\bf F}_{m} \phi_{m}, & \\
	&  \vartheta_{m}'' + \frac{1}{r}\vartheta_{m}' - \left(k^{2} + 
    \frac{m^{2}+1}{r^{2}} \right) \vartheta_{m} - \frac{2m}{r^{2}} 
    \varrho_{m}  = & \nonumber \\
    &   - \frac{m}{\eta r} P_{m}. &
\end{eqnarray}
\end{mathletters}
Adding these together we find
\begin{eqnarray*}
	& (\varrho_{m}+\vartheta_{m})'' + 
	\frac{1}{r}(\varrho_{m}+\vartheta_{m})' - 
	\left(k^{2} + \frac{(m+1)^{2}}{r^{2}}\right) (\varrho_{m} + 
	\vartheta_{m}) = &  \nonumber \\
	&  \frac{1}{\eta} P_{m}' - \frac{m}{\eta r} P_{m} - 
	\frac{1}{\eta} \phi_{s}' {\bf F}_{m} \phi_{m}. &
\end{eqnarray*}
The homogeneous part of this equation is the modified Bessel equation, 
with general solutions $I_{m+1}(kr)$ and $K_{m+1}(kr)$. Similarly 
subtracting gives
\begin{eqnarray*}
	& (\varrho_{m}-\vartheta_{m})'' + 
	\frac{1}{r}(\varrho_{m}-\vartheta_{m})' - 
	\left(k^{2} + \frac{(m-1)^{2}}{r^{2}}\right) (\varrho_{m} - 
	\vartheta_{m}) & \nonumber \\
	&  = \frac{1}{\eta} P_{m}' + \frac{m}{\eta r} P_{m} - 
	\frac{1}{\eta} \phi_{s}' {\bf F}_{m} \phi_{m}, &
\end{eqnarray*}
so now the homogeneous part of the equation has solutions 
$I_{m-1}(kr)$ and $K_{m-1}(kr)$.  We can thus construct exact Green's 
functions for the combinations $\varrho_{m}+\vartheta_{m}$ and 
$\varrho_{m}-\vartheta_{m}$. The solutions to the inhomogeneous equations 
are therefore

\end{multicols}
\begin{picture}(3.5,.25)
\put(0,.125){\line(1,0){3}}
\end{picture}

\begin{mathletters}
\label{vsoltn}
\begin{eqnarray}
	\varrho_{m}(r)+\vartheta_{m}(r) & = & - \int_{0}^{\infty} r'\, dr' 
	\, K_{m+1}(kr_{>})I_{m+1}(kr_{<}) \left(\frac{1}{\eta}P_{m}'(r') -
	\frac{m}{\eta r'} P_{m}(r') - \frac{1}{\eta} \phi_{s}' {\bf F}_{m} 
	\phi_{m}(r') \right), \\
	\varrho_{m}(r)-\vartheta_{m}(r)  & = &  - \int_{0}^{\infty} r'\, dr' \, 
	K_{m-1}(kr_{>})I_{m-1}(kr_{<})  \left(\frac{1}{\eta} P_{m}'(r') +
	\frac{m}{\eta r'} P_{m}(r') - \frac{1}{\eta} \phi_{s}' {\bf F}_{m} 
	\phi_{m}(r') \right).
\end{eqnarray}
\end{mathletters}
All of the functions in these integrals are known, so we have reduced 
the solution for $v_{r,m}(r)$ to quadrature. For the case of 
equal viscosities, we could therefore again integrate numerically 
to find the matrix elements for finite width interfaces.

Instead I will proceed with the sharp interface approximation Eq.\ 
(\ref{sharpint}).  In this limit the velocity matrix elements in Eq.\ 
(\ref{matrix1}), divided by the normalization integrals, are
\begin{equation}
	\omega_{m,h}  \cong
	\frac{1}{(\tilde{\phi}_{m},\phi_{m})} \int_{0}^{\infty} dr\, r 
	\tilde{\phi}_{m}(r) v_{r,m}(r) 2\delta(r-R) 
	 =  v_{r,m}(R). \label{wh}
\end{equation}
Thus $\omega_{m,h}$ is given not surprisingly by the radial 
velocity at the linearized position of the interface, $v_{r,m}$ 
evaluated at $r=R$.  From Eq.\ (\ref{pressure}) we find the 
pressure $P_{m}$ is given by
\begin{equation}
	P_{m}(r) = \frac{2(k^{2}R^{2}+m^{2}-1)}{3R} \left\{ 
	\begin{array}{cc} kK_{m}'(kR)I_{m}(kr), & r<R \\
	kK_{m}(kr)I_{m}'(kR), & r>R. \end{array} \right. 
\end{equation}
Substituting into Eq.\ (\ref{vsoltn}), solving for $v_{r}$, and performing 
some reductions using Bessel function identities, we find
\begin{eqnarray}
	\omega_{m,h} & = & \frac{2(k^{2}R^{2}+m^{2}-1)}{3\eta 
	R}\left\{\frac{1}{2}(m+1) K_{m+1}(q)I_{m+1}(q) - \frac{1}{2}(m-1) 
	K_{m-1}(q)I_{m-1}(q) \right. \nonumber \\
	& &  + \frac{1}{2}q[K_{m}(q)I_{m+1}(q) - 
	K_{m-1}(q)I_{m}(q)] \nonumber \\
	& & \left. + \frac{1}{8}q^{2}[K_{m+1}(q)I_{m+1}(q) - 
	K_{m-1}(q)I_{m-1}(q)] [K_{m-1}(q)I_{m+1}(q)-K_{m+1}(q)I_{m-1}(q)] \right\}
	\label{vrm}
\end{eqnarray}

\begin{picture}(6.5,.25)
\put(3.75,.125){\line(1,0){3}}
\end{picture}

\begin{multicols}{2}

where we have set $q \equiv kR$ for convenience.  These give the 
hydrodynamic part of the stability eigenvalues in the absence of the 
shear flow.  Since I have taken the sharp interface limit, this part 
of the calculation has been completely decoupled from the dynamics of 
the concentration field $\phi$; Eq.\ (\ref{wh}) is equivalent to the 
usual kinematic condition in hydrodynamic stability calculations.  
Thus the result Eq.\ (\ref{vrm}) is an exact, rather than approximate, 
solution for the hydrodynamic stability of a cylinder of one fluid 
immersed in a second immiscible one.  It can therefore be compared 
directly with previous results.  For the varicose mode $m=0$ we obtain 
the stability eigenvalue
\begin{eqnarray}
	\omega_{0,h}(k) & = & \frac{2(k^{2}R^{2}-1)}{3\eta 
	R}\left[K_{1}(q)I_{1}(q) + \frac{1}{2} qK_{0}(q)I_{1}(q) \right. 
	\nonumber \\
	& & \left. - \frac{1}{2}qK_{1}(q)I_{0}(q) \right].
\end{eqnarray}
Putting this back in dimensional form we have
\begin{eqnarray}
	\varpi_{0,h}(k) & = & \frac{\sigma (q^{2}-1)}{\nu 
	{\cal R}}\left[K_{1}(q)I_{1}(q) + \frac{1}{2} qK_{0}(q)I_{1}(q) 
	\right.\nonumber \\
	& & \left. - \frac{1}{2}qK_{1}(q)I_{0}(q) \right].
	\label{tomotika}
\end{eqnarray}
This expression is the same as that found by Stone and Brenner 
\cite{stone} and may be obtained from Tomotika's general result for 
the dispersion relation of the viscous Rayleigh instability, in the 
limit of equal viscosities between the two liquids \cite{tomotika}.

\begin{figure}
\begin{center}
\leavevmode
\epsfxsize= 4in
\epsfbox{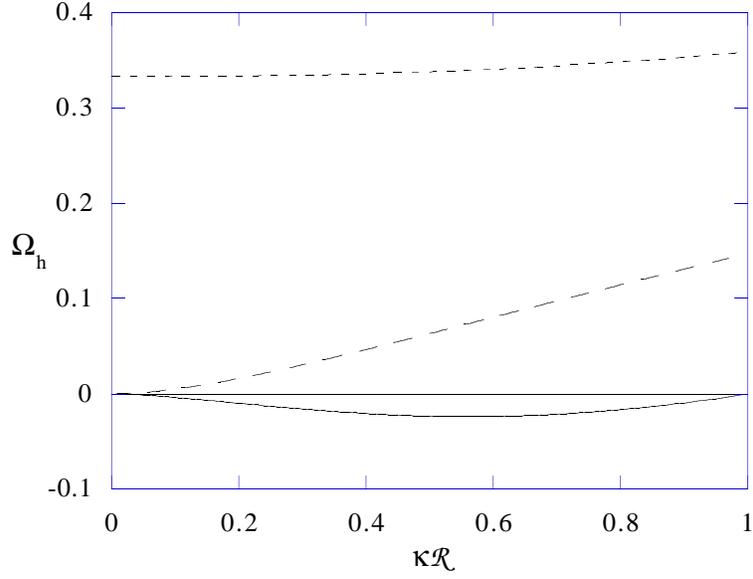}
\end{center}
\caption{\label{disp0} Stability eigenvalues $\Omega_{0,h}$ (solid 
curve), $\Omega_{1,h}$ (dashed curve), and $\Omega_{2,h}$ (dotted curve) 
for $\mu=1$ in the absence of shear.}
\end{figure}

Note each mode in Eq.\ (\ref{vrm}) can be written in the original 
variables as $\varpi_{m,h} = \sigma \Omega_{m,h}(\kappa {\cal R})/\nu 
{\cal R}$, where $\Omega_{m,h}$ is dimensionless.  This dimensionless 
part of the dispersion relation is graphed in Fig.\ \ref{disp0} for 
the first three modes.  Again the varicose mode is unstable for all 
$\kappa<1/{\cal R}$, whereas the higher modes are stable for all 
$\kappa$.

\subsection{Velocity field, general viscosity ratio}
\label{vmu}

For the case of general viscosity ratio $\mu$, it is much more 
difficult to solve the hydrodynamic equation with diffuse 
interfaces.  To do so requires writing the viscosity as a smooth 
function of $\phi$ that changes from the value inside the cylinder 
$\eta^{i}$ to the value outside $\eta^{o}$ in a continuous way, 
which introduces extra terms into the original differential 
equation (\ref{N-S}).  Thus in this section I will start with 
sharp interfaces from the beginning.  It is then more sensible to 
follow a different approach to solving the hydrodynamic equation.  
For sharp interfaces, the term coupling the total concentration 
$\Phi$ to the total velocity ${\bf u}$ in Eq.\ (\ref{N-S}) becomes 
\cite{chella}
\begin{displaymath}
	\nabla \Phi \frac{\delta F}{\delta \Phi} \rightarrow \sigma 
	h \delta(\zeta) {\bf \hat{n}},
\end{displaymath}
where $h$ is the curvature of the interface located at $\zeta({\bf 
r})=0$ and ${\bf \hat{n}}$ is a unit vector normal to the interface.  
This is equivalent to the usual boundary condition on the jump in the 
normal stress across a fluid interface.  Instead of including this 
coupling term in the hydrodynamic equation, we can just solve the 
usual creeping flow equations for the perturbed velocity field ${\bf 
v}^{i}$ inside and ${\bf v}^{o}$ outside the cylinder separately, and 
apply the appropriate boundary conditions at the interface in the 
usual manner.  In each region ${\bf v}$ satisfies
\begin{equation}
	\nabla^{2} {\bf v} = \frac{1}{\eta} {\bf \nabla}P, \quad 0 = {\bf 
	\nabla} \cdot {\bf v},
	\label{hydro3}
\end{equation}
and the pressure simply satisfies the Laplace equation,
\begin{equation}
	\nabla^{2} P = 0.
	\label{peq}
\end{equation}
The location of the interface of course depends on the perturbation 
mode $\phi_{m}$ in question; since $\phi_{m}(r)$ is now a step 
function we only need know the location of the step, which is given by
\begin{equation}
	r = R - \epsilon e^{im\theta + ikx}
\end{equation}
for each mode $m$, where $\epsilon$ is the amplitude of the small 
perturbation. 

The solutions to Eqs.\ (\ref{hydro3})--(\ref{peq}) are 
straightforward so the details are given in the appendix.  The 
general solutions are found in terms of modified Bessel functions; 
the exact solutions are then found by applying the boundary 
conditions and solving numerically for the unknown constants.  The 
results are the coefficients ${\bf v}_{nm}(r)$ in the expansion 
(\ref{vnm}).  As shown in the appendix, the only nonzero 
coefficients are the ones for which $n=m$ and $n=m \pm 1$ so Eq.\ 
(\ref{matrix1}) is tridiagonal as written.  As was shown in the 
last section in Eq.\ (\ref{wh}), the velocity matrix elements are 
then simply given by
\begin{equation}
	\frac{(\tilde{\phi}_{n}, v_{r,nm}(r)\phi_{s}')} 
	{(\tilde{\phi}_{n}, \phi_{m})} = v_{r,nm}(r=R).
\end{equation}
In the case of zero shear flow, only the diagonal matrix elements are 
nonzero and the hydrodynamic part of the stability eigenvalues is 
therefore $\omega_{m,h} = v_{r,mm}(R)$.  Again the results here for 
$S=0$ are decoupled from $\phi$ and are therefore exact, for two 
immiscible liquids.

\begin{figure}
\begin{center}
\leavevmode
\epsfxsize= 4in
\epsfbox{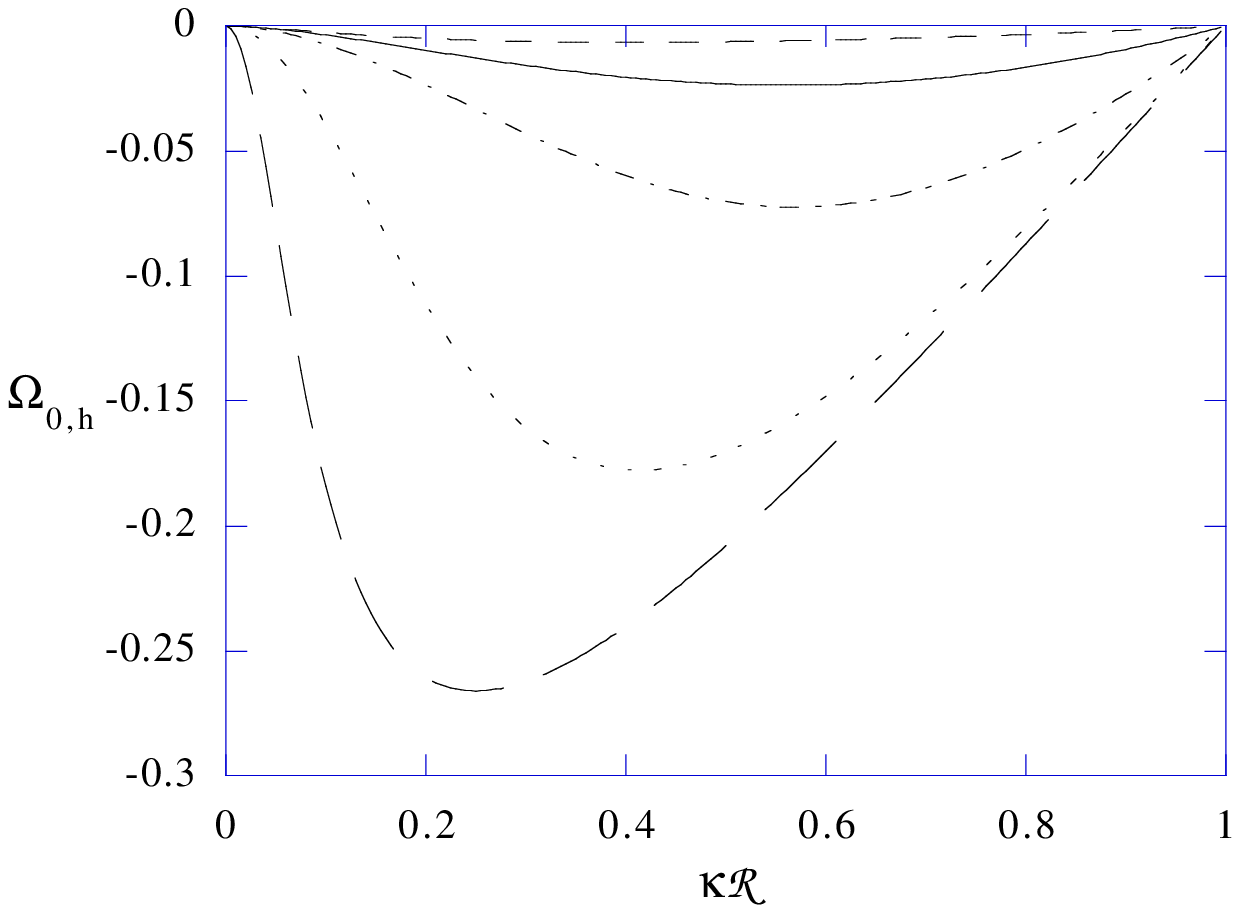}
\end{center}
\caption{\label{w0hmu} Stability eigenvalue $\Omega_{0,h}$ 
for $S=0$ at different $\mu$: $\mu=10$ (dashed curve), 
$\mu=1$ (solid curve), $\mu=.1$ (dash-dot curve), $\mu=.01$ (dotted 
curve), and $\mu=.001$ (long dashed curve).}
\end{figure}

\begin{figure}
\begin{center}
\leavevmode
\epsfxsize= 4in
\epsfbox{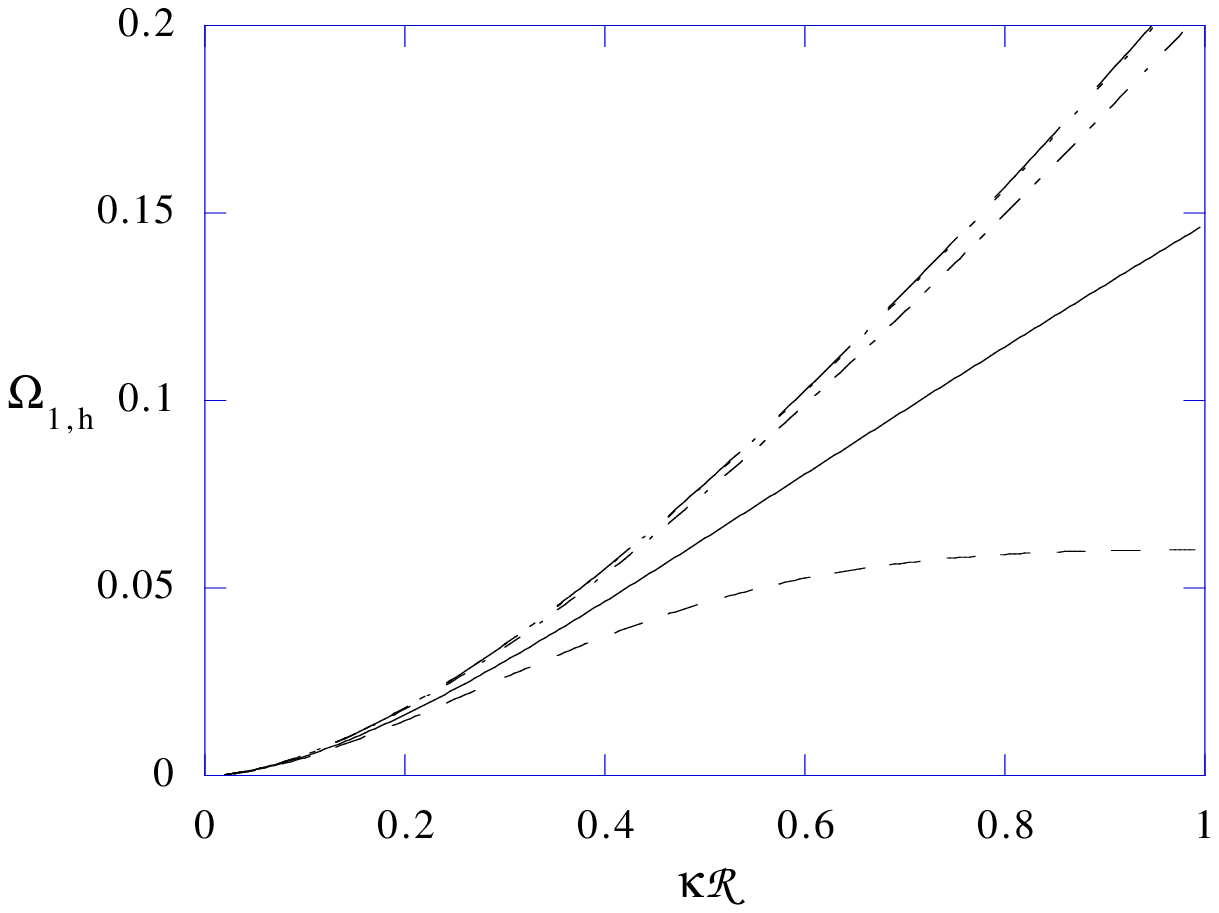}
\end{center}
\caption{\label{w1hmu} Stability eigenvalue
$\Omega_{1,h}$ for $S=0$ at different $\mu$: $\mu=10$ (dashed curve), 
$\mu=1$ (solid curve), $\mu=.1$ (dash-dot curve), $\mu=.01$ (dotted 
curve), and $\mu=.001$ (long dashed curve).}
\end{figure}

\begin{figure}
\begin{center}
\leavevmode
\epsfxsize= 4in
\epsfbox{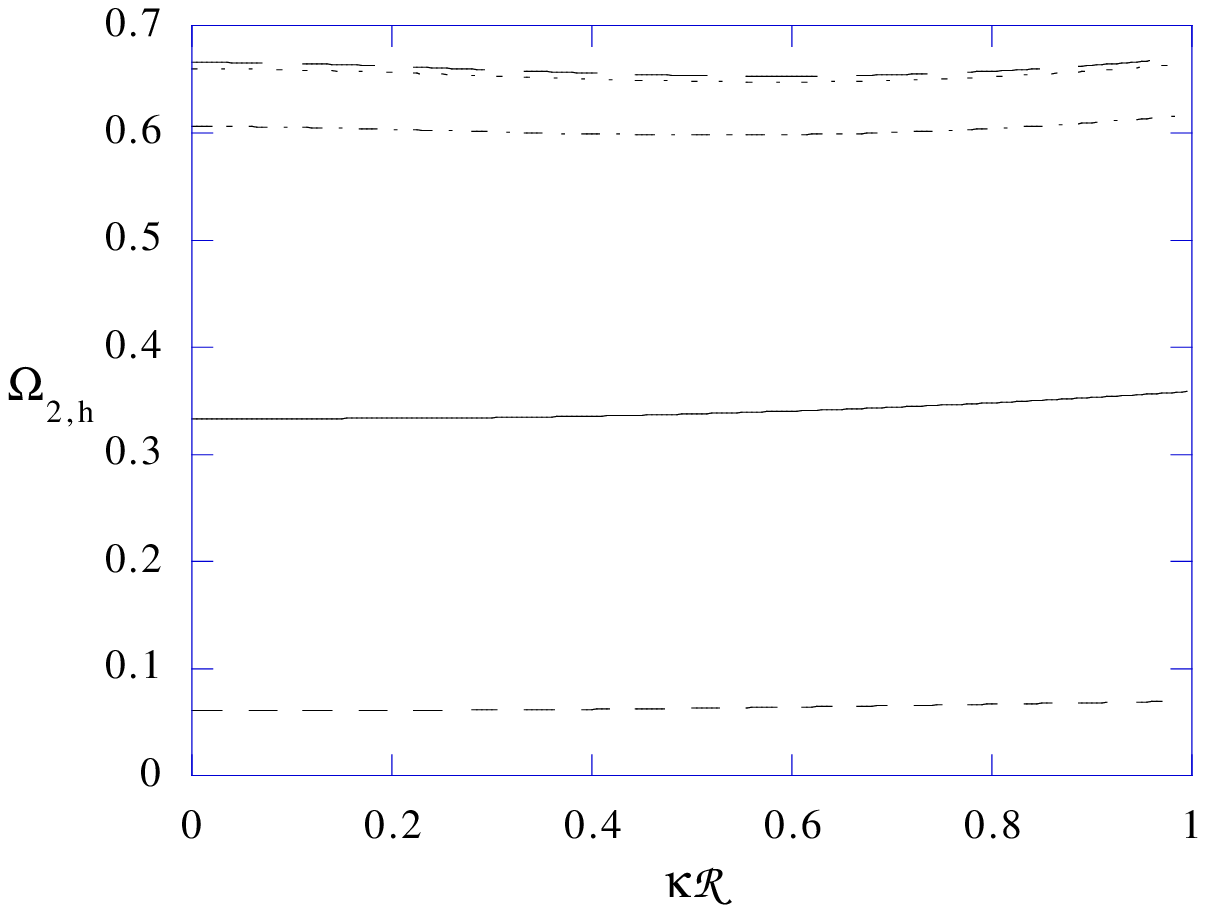}
\end{center}
\caption{\label{w2hmu} Stability eigenvalue $\Omega_{2,h}$ for $S=0$ 
at different $\mu$: $\mu=10$ (dashed curve), $\mu=1$ (solid curve), 
$\mu=.1$ (dash-dot curve), $\mu=.01$ (dotted curve), and $\mu=.001$ 
(long dashed curve).}
\end{figure}

I show in the appendix that in the absence of the shear flow the 
velocity matrix elements and hence the eigenvalues are proportional to 
$1/\eta^{o}R$, just as in the $\mu=1$ case.  Figs.\ 
\ref{w0hmu}--\ref{w2hmu} show the dispersion relations for the 
dimensionless parts of the three lowest modes for different values of 
$\mu$ (holding $\eta^{o}$ constant).  The results for the varicose 
$m=0$ mode are again the same as those of Tomotika \cite{tomotika}.  
As the cylindrical domain becomes less viscous relative to the 
background, it becomes more unstable and the wave number of maximum 
instability $\kappa_{max}$ becomes smaller.  Note that the damping 
rate for the undulatory $m=1$ mode depends less strongly on $\mu$.  
From Fig.\ \ref{w2hmu} we see that the $m=2$ depends only weakly on 
$\kappa{\cal R}$.

Finally, in the appendix the hydrodynamic equations are solved 
analytically at $k=0$, providing a check on the numerical results.  
The stability eigenvalues at $k=0$ are diagonal and independent of the 
shear rate $S$, Eq.\ (\ref{k0exact}):
\begin{equation}
	\omega_{m,h}(k=0) = \frac{m}{3\eta^{o} R (\mu +1)}, \quad m \geq 2,
	\label{wk0}
\end{equation}
with $\omega_{0,h}(k=0) = \omega_{1,h}(k=0) = 0$.  Note that since the 
only remaining off-diagonal elements in Eq.\ (\ref{matrix1}) vanish at 
$k=0$, this demonstrates that the cylinder is stable towards 
$x$-independent deformations which would result in a non-circular 
cross-section, even in the presence of the shear flow.

\subsection{Shear flow contribution}

Finally we evaluate the off-diagonal shear flow matrix elements 
\begin{displaymath}
	\frac{1}{2} ik \frac{(\tilde{\phi}_{n},u_{s}(r) 
	\phi_{m})}{(\tilde{\phi}_{n}, \phi_{m})},
\end{displaymath}
where we have divided by the normalization integral.  Since we are 
assuming that $\phi_{m}(r)$ has the same radial dependence for all 
$m$, we simply have in the sharp interface limit
\begin{eqnarray}
	\frac{ik}{2} \frac{(\tilde{\phi}_{n},u_{s}(r)
	\phi_{m})}{(\tilde{\phi}_{n}, \phi_{m})} & = &
	\frac{ik}{2(\tilde{\phi}_{n}, \phi_{m})} 
	\int_{0}^{\infty} dr \, r u_{s}(r) \tilde{\phi}_{n} \phi_{m} \nonumber \\
	 & \cong & \frac{ik}{2(\tilde{\phi}_{n}, \phi_{m})} 
	 \int_{0}^{\infty} dr \, ru_{s}(r) \tilde{\phi}_{n} 2 \delta(r-R) 
	 \nonumber \\
	 & = &   \frac{1}{2} iku_{s}(R) = \frac{ikRS}{\mu+1}.
	\label{shear}
\end{eqnarray}
Thus each off-diagonal element is the same independent of $m$.

\section{Results with shear}
\label{shearres}

With all the necessary matrix elements in hand, we can now solve 
for the stability eigenvalues in shear by diagonalizing Eq.\ 
(\ref{matrix1}).  To do so requires truncating the matrix at some 
point.  Since the off-diagonal matrix elements are all 
proportional to $S$, we can assume that any perturbation which 
damps out more quickly than the time associated with the shear 
flow can be ignored.  Thus we only need include modes whose 
damping rates are less than or of the order of the applied shear 
rate.  For comparison purposes, the values of the stability 
eigenvalues at $k=0$ give good estimates of the damping rates of 
high $m$ modes without having to calculate the full dispersion 
relations.

In this paper it will be sufficient to only include the first 
three modes, in which case the matrix equation (\ref{matrix1}) 
becomes a $3\times 3$ secular equation for $\omega$:
\begin{equation}
	\left | \begin{array}{ccc} \omega_{0}-\omega & \frac{ikRS}{\mu+1} + 
	iS\omega_{01} & 0 \\
	\frac{ikRS}{\mu+1} + iS\omega_{10} & \omega_{1}-\omega & 
	\frac{ikRS}{\mu+1} + iS\omega_{12} \\
	0 & \frac{ikRS}{\mu+1} + iS \omega_{21} & \omega_{2}-\omega \end{array} 
	\right | = 0,
	\label{matrix2}
\end{equation}
where I have pulled a factor of $iS$ out of the off-diagonal velocity 
matrix elements, $v_{nm,r}(R) \equiv iS\omega_{nm}$.  This leads to a 
cubic characteristic equation for the stability eigenvalues 
$\omega(k)$.  Since the characteristic equation has real coefficients, 
the roots can be found analytically; there will be either three real 
roots or one real root and a complex conjugate pair.

Recall that in the absence of the shear flow, the stability 
eigenvalues are simply given by the sum of the diffusive and 
hydrodynamic terms as in Eq.\ (\ref{evm}):
\begin{displaymath}
	\omega_{m} = \omega_{m,d} + \omega_{m,h}, \quad S = 0 \,.
\end{displaymath}
I showed in Section \ref{diffusive} and Section \ref{vmu} that 
these two terms scale differently with the parameters of the 
system, with $\omega_{m,d} = \Omega_{m,d}(kR)/R^{3}$ and 
$\omega_{m,h} = \Omega_{m,h}(kR,\mu)/\eta^{o}R$.  The relative 
magnitude of these two terms depends on the dimensionless 
viscosity parameter $\eta^{o} = 2\xi\nu^{o}/3\sigma\tau_{\xi}$ and 
on the radius of the cylinder.  For sufficiently viscous and/or 
small cylindrical domains, the diffusive term will dominate, 
whereas for less viscous and/or large cylindrical domains, the 
hydrodynamic term dominates.  In the following I first examine the 
stability eigenvalues in the two extreme limits in which the 
hydrodynamic terms or the diffusive terms can be disregarded 
entirely.  I will then present some results for experimentally 
realistic parameter values.

\subsection{Stabilization of the Rayleigh instability}
\label{Rstable}

First consider Eq.\ (\ref{matrix2}) in the limit that the diffusive 
terms in the diagonal elements are negligible, so that $\omega_{m} 
= \omega_{m,h}$.  Physically this corresponds to examing the effect of 
the shear flow on the purely hydrodynamic Rayleigh instability.  Since 
the diagonal matrix elements $\omega_{m,h}$ scale as $1/\eta^{o}R$ 
while the off-diagonal elements do not depend on $\eta^{o}$ or $R$, 
the stability eigenvalues $\omega$ will also scale with $1/\eta^{o}R$.  
Denote the eigenvalues found by diagonalizing Eq.\ (\ref{matrix2}) 
with the shear flow present by
\begin{equation}
	\omega_{j,h}(S) = \frac{1}{\eta^{o}R} \Omega_{j,h}(kR,\mu,S),
\end{equation}
where the index $j$ refers to the order of the new eigenvalues: 
$\omega_{0}(S)<\omega_{1}(S)<\ldots$.  Returning to the original 
variables, this relation is simply
\begin{equation}
	\varpi_{j,h}(S) = \frac{\sigma}{\nu^{o}\cal{R}}
	\Omega_{j,h}(kR,\mu,S).
\end{equation}
Since $\sigma/\nu^{o}{\cal R}$ has units of inverse time, we can 
measure the shear rate $\dot{\gamma}$ just as well in these units as 
in units of $1/\tau_{\xi}$, so for the rest of this section I will 
take $S \equiv \nu^{0}{\cal R} \dot{\gamma}/\sigma$.

\begin{figure}
\begin{center}
\leavevmode
\epsfxsize= 4in
\epsfbox{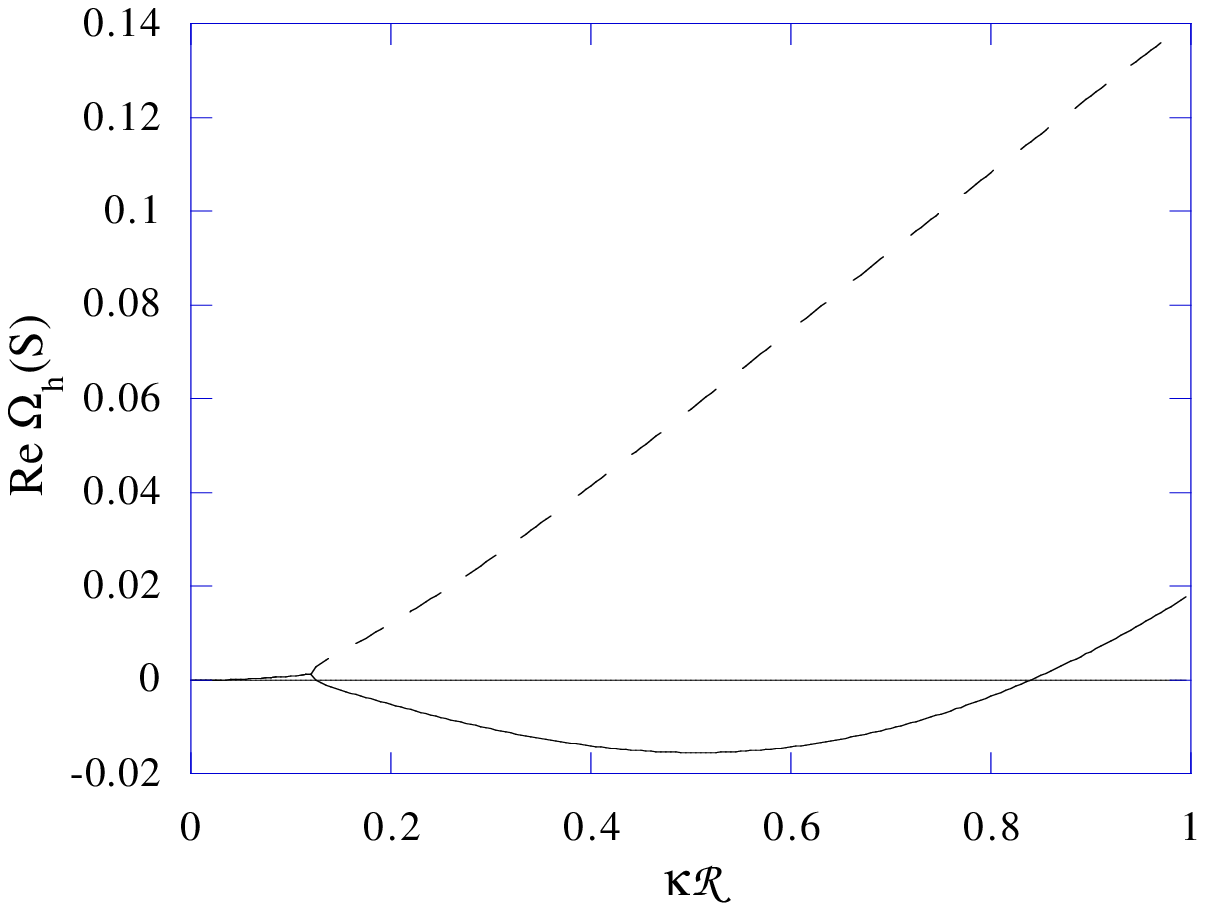}
\end{center}
\caption{\label{mu1gp1} Stability eigenvalues $Re[\Omega_{0,h}(S)]$ 
(solid line) and $Re[\Omega_{1,h}(S)]$ (dashed line) for $\mu=1$ at $S = 
.1$.}
\end{figure}

\begin{figure}
\begin{center}
\leavevmode
\epsfxsize= 4in
\epsfbox{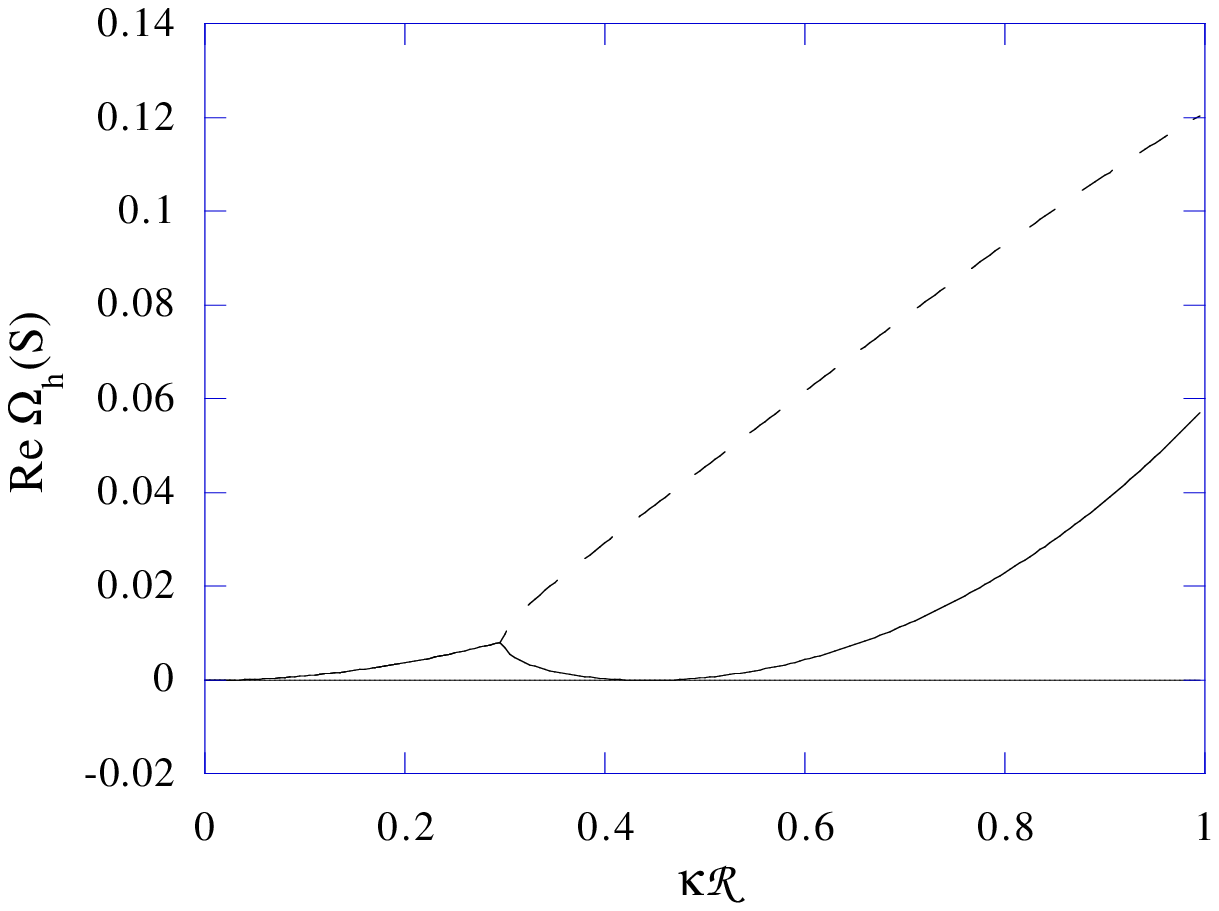}
\end{center}
\caption{\label{mu1gc} Stability eigenvalues $Re[\Omega_{0,h}(S)]$ 
(solid line) and $Re[\Omega_{1,h}(S)]$ (dashed line) for $\mu=1$ at the 
critical shear rate $S_{c} = .160$.}
\end{figure}

\begin{figure}
\begin{center}
\leavevmode
\epsfxsize= 4in
\epsfbox{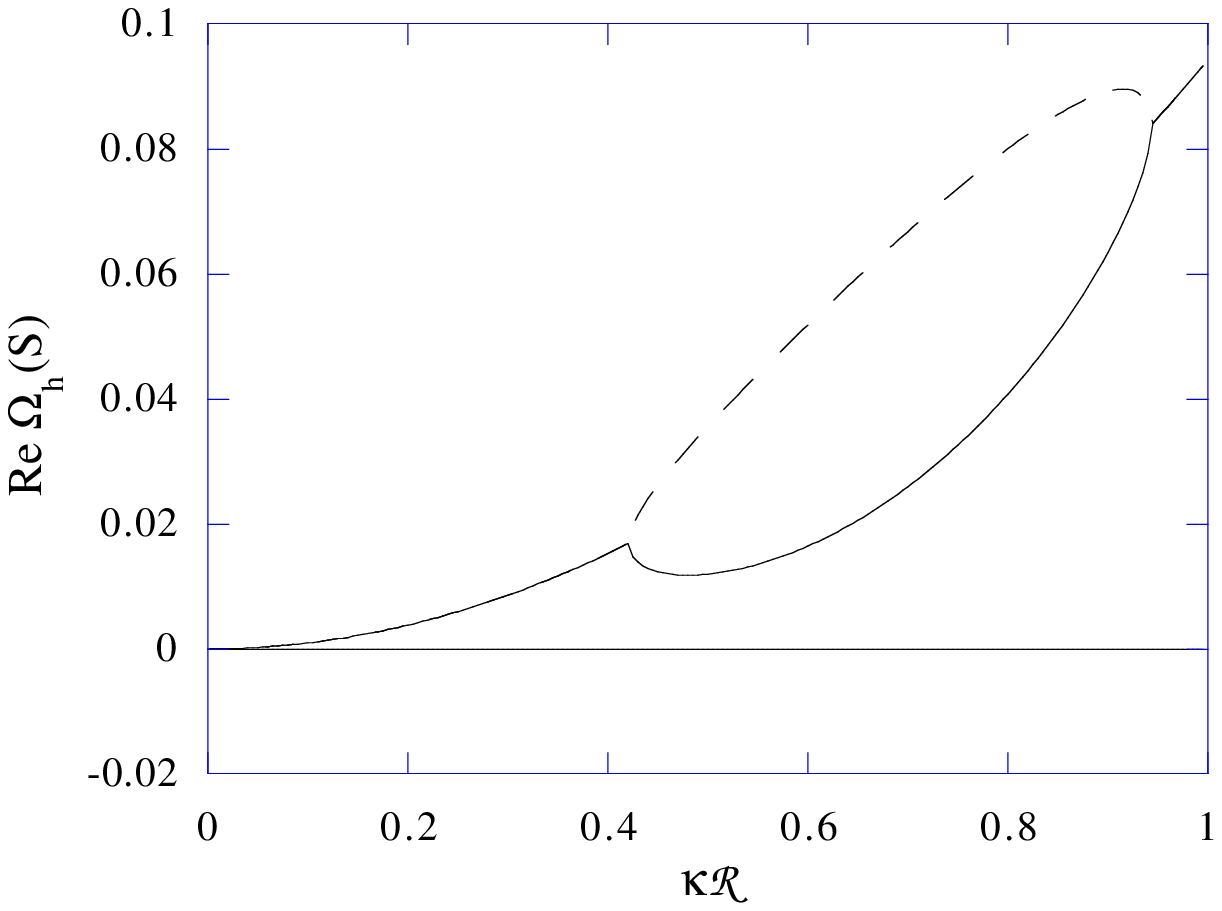}
\end{center}
\caption{\label{mu1gp18} Stability eigenvalues $Re[\Omega_{0,h}(S)]$ 
(solid line) and $Re[\Omega_{1,h}(S)]$ (dashed line) for $\mu=1$ at $S = 
.18$.}
\end{figure}

We start by considering the equal viscosity case, $\mu=1$.  Figs.\ 
(\ref{mu1gp1})--(\ref{mu1gp18}) show the real part of the 
dimensionless dispersion relations $\Omega_{j,h}(kR,1,S)$ for the 
first two stability eigenvalues in shear flow at various shear rates 
(the third mode was included in the calculation but is not a 
particularly interesting function of $S$).  For some values of 
$\kappa$ the lowest two modes $\omega_{0,h}(S)$ and $\omega_{1,h}(S)$ 
are a complex conjugate pair and so they show up as a single curve in 
Figs.\ (\ref{mu1gp1})--(\ref{mu1gp18}); for the regions in $\kappa$ 
for which there are two curves shown, the two eigenvalues are both 
completely real.  We see that at small $\kappa$ where there 
is a complex conjugate pair, the perturbations are traveling waves 
with an overall damping rate.  At low shear rates the mode with 
the lowest eigenvalue $\omega_{0,h}(S)$ is still unstable, but the 
window of wave numbers over which $\omega_{0,h}(S)<0$ becomes smaller 
as $S$ increases.  At some critical shear rate $S_{c}$ the minimum in 
$\omega_{0,h}(S)$ crosses zero, at a critical wave number 
$\kappa_{c}=k_{c}/{\cal R}$.  Above this critical shear rate, the 
instability is gone---the initially unstable varicose mode has been 
stabilized by the applied shear flow, by being mixed with the higher 
modes.  

Clearly, the critical shear rate $\dot{\gamma}_{c}$ must have the same 
dependence on $\nu^{o}$, $\sigma$, and $\cal{R}$ as the stability 
eigenvalues:
\begin{equation}
	\dot{\gamma}_{c} = \frac{\sigma S_{c}(\mu)}{\nu^{o}\cal{R}}.
	\label{gammac}
\end{equation}
Thus the critical shear rate is a monotonically decreasing function of 
the radius of the cylinder and the magnitude of the outside viscosity; 
the smaller the cylinder or the less viscous the fluids are overall, 
the faster the growth of the varicose mode and so the critical shear 
rate must be faster as well for stabilization to occur.  Inverting Eq.\ 
(\ref{gammac}) we see that at a fixed shear rate, there is a 
critical radius $R_{c} = \dot{\gamma} \nu^{o}/\sigma S(\mu)$ above 
which the cylinder is stable and below which it is unstable.  Thus if 
instead of an infinite cylinder we had a finite long cylindrical drop 
that was being stretched by the flow, initially small capillary 
disturbances on the drop would be suppressed by the flow, but as the 
drop thinned to a radius smaller than $R_{c}$ the disturbances would 
start to grow and the drop would break up. Note that $S_{c}$ is 
considerably larger than the magnitudes of both $\Omega_{0,h}(S)$ and 
$\Omega_{1,h}(S)$ at all $\kappa {\cal R}<1$, so the shear rate must 
be faster than the rate of growth of the instability to stabilize it.

\begin{figure}
\begin{center}
\leavevmode
\epsfxsize= 4in
\epsfbox{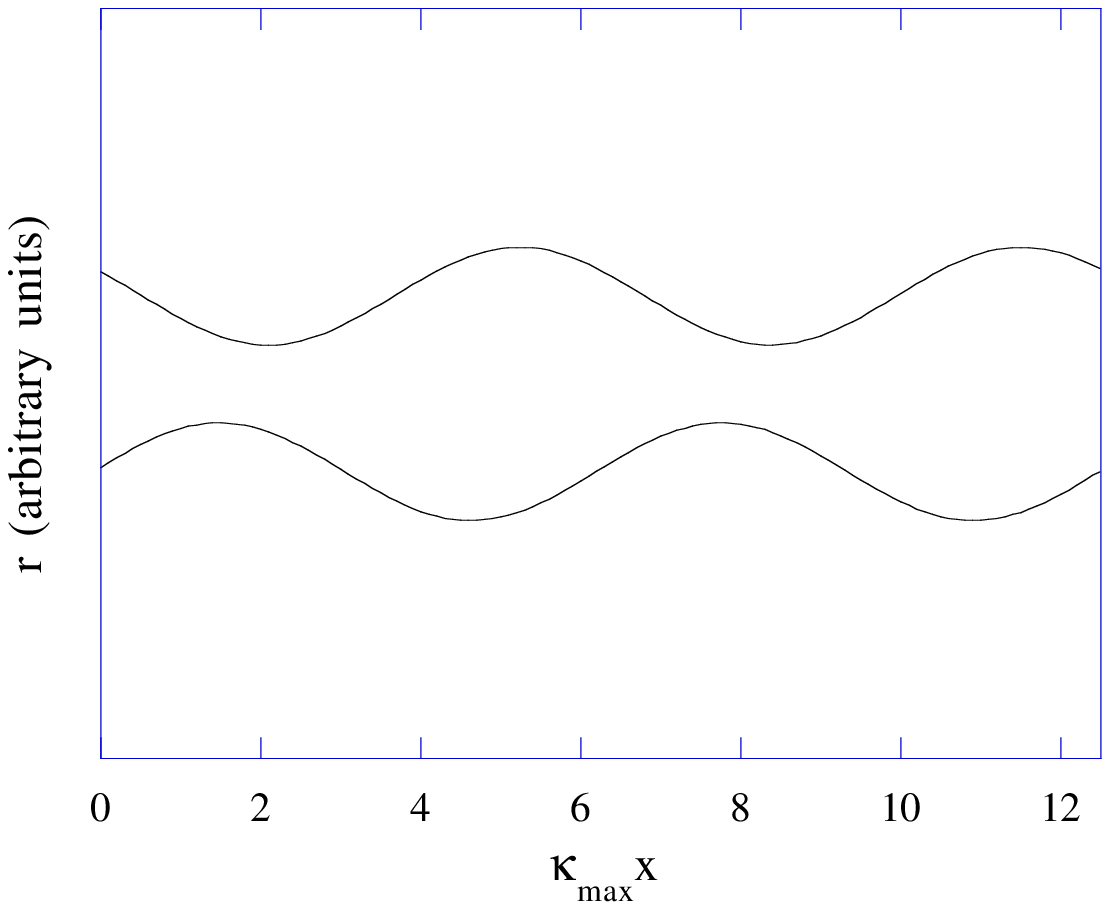}
\end{center}
\caption{\label{evSgp1} Real part of the eigenvector corresponding to 
$\Omega_{0,h}$ at $S = .1$, in the $\theta=0$ plane.}
\end{figure}

\begin{figure}
\begin{center}
\leavevmode
\epsfxsize= 4in
\epsfbox{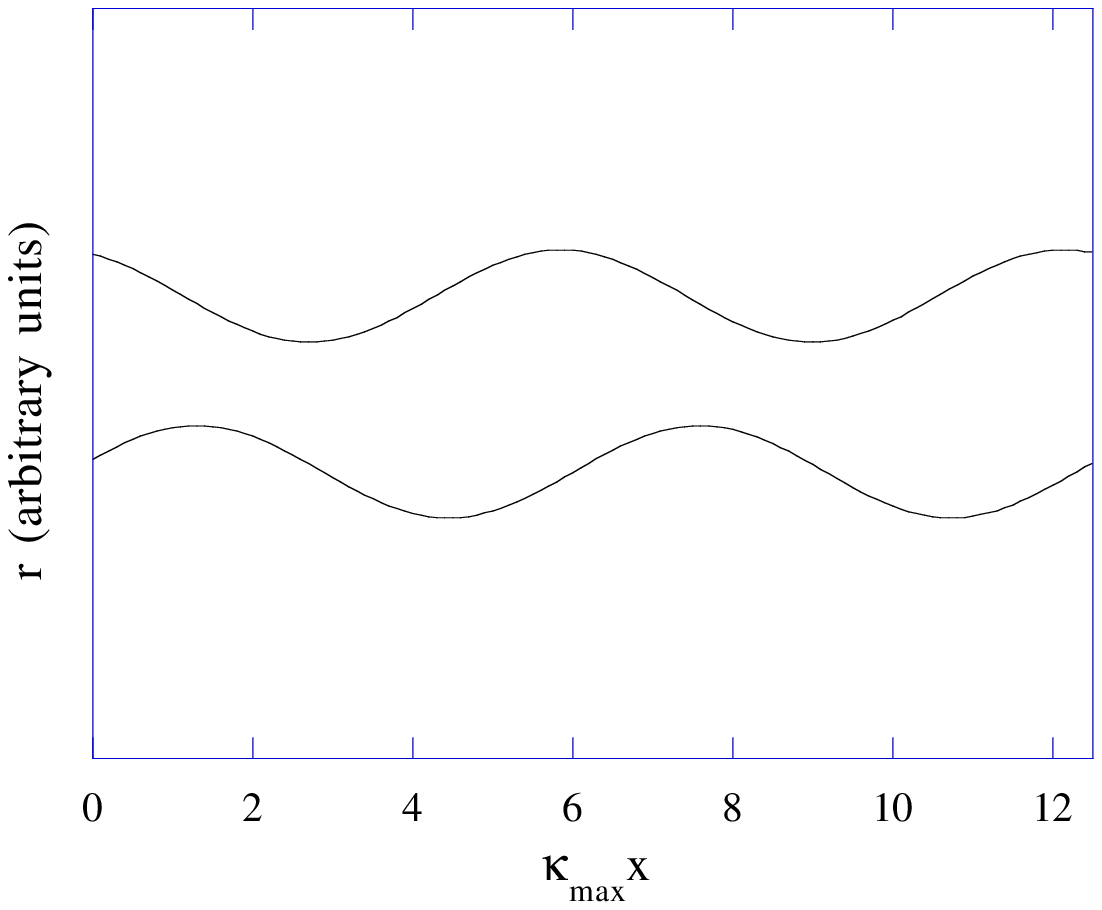}
\end{center}
\caption{\label{evSgp18} Real part of the eigenvector corresponding to 
$\Omega_{0,h}$ at $S = .18$, in the $\theta=0$ plane.}
\end{figure}

The qualitative picture here is that the shear flow advects opposite 
sides of the cylinder relative to each other so that the special 
axisymmetric, varicose perturbation no longer exists long enough to be 
unstable.  This picture is born out by the eigenvectors corresponding 
to the eigenvalues shown in Figs.\ \ref{mu1gp1}--\ref{mu1gp18}.  Fig.\ 
\ref{evSgp1} shows a cross-section of the eigenvector corresponding to 
the lowest eigenvalue at $\kappa_{max} {\cal R}=.56219$ (where 
$\kappa_{max}$ is the wave number of maximum instability in zero 
shear) and $S=.1$, when the lowest mode is still unstable; Fig.\ 
\ref{evSgp18} shows the same eigenvector at $S=.18$, which is above 
the critical shear rate and therefore stable.  As the shear rate 
increases, the original varicose mode becomes more distorted as it is 
mixed with the other modes.

\begin{figure}
\begin{center}
\leavevmode
\epsfxsize= 4in
\epsfbox{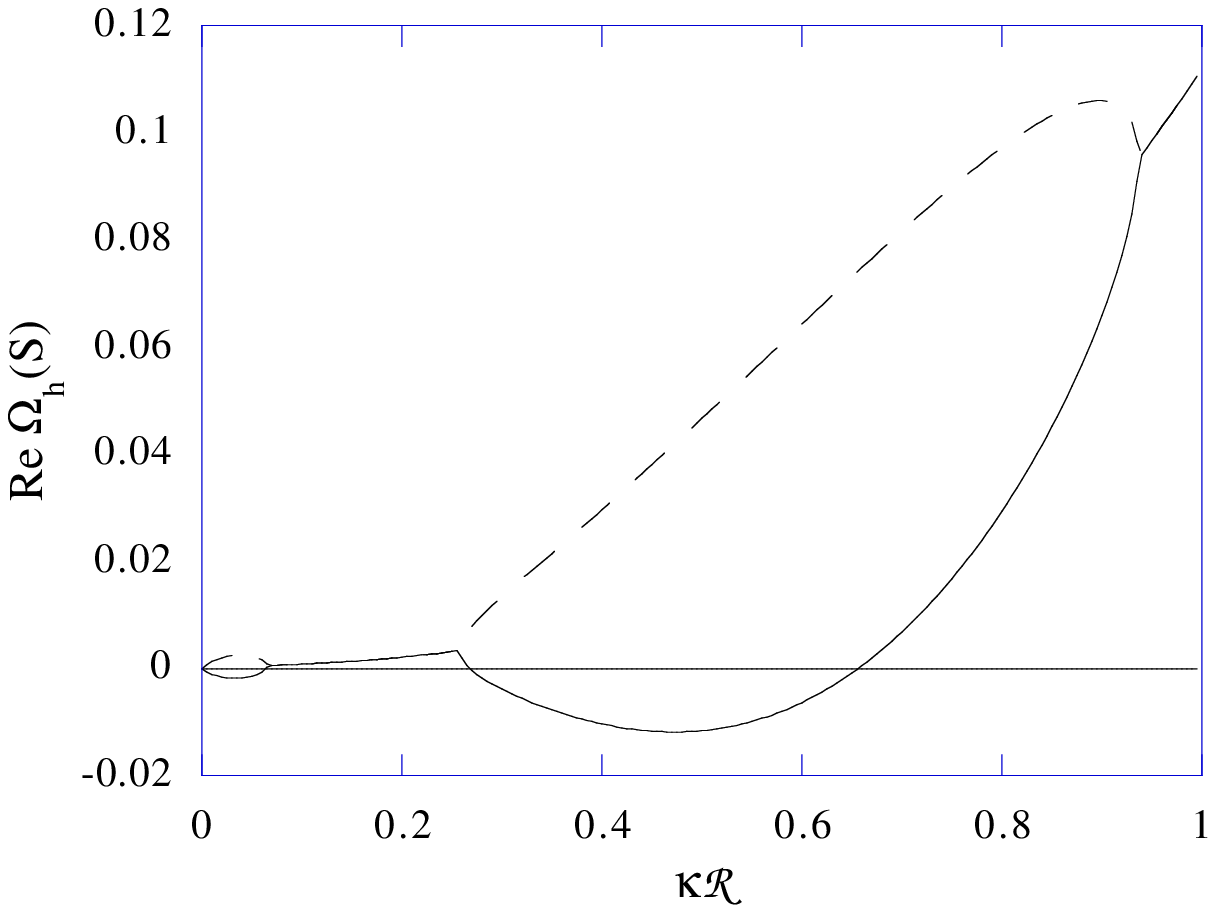}
\end{center}
\caption{\label{mup25gp14} Stability eigenvalues $Re[\Omega_{0,h}(S)]$ 
(solid line) and $Re[\Omega_{1,h}(S)]$ (dashed line) for $\mu=.25$ at $S = 
.14$.}
\end{figure}

\begin{figure}
\begin{center}
\leavevmode
\epsfxsize= 4in
\epsfbox{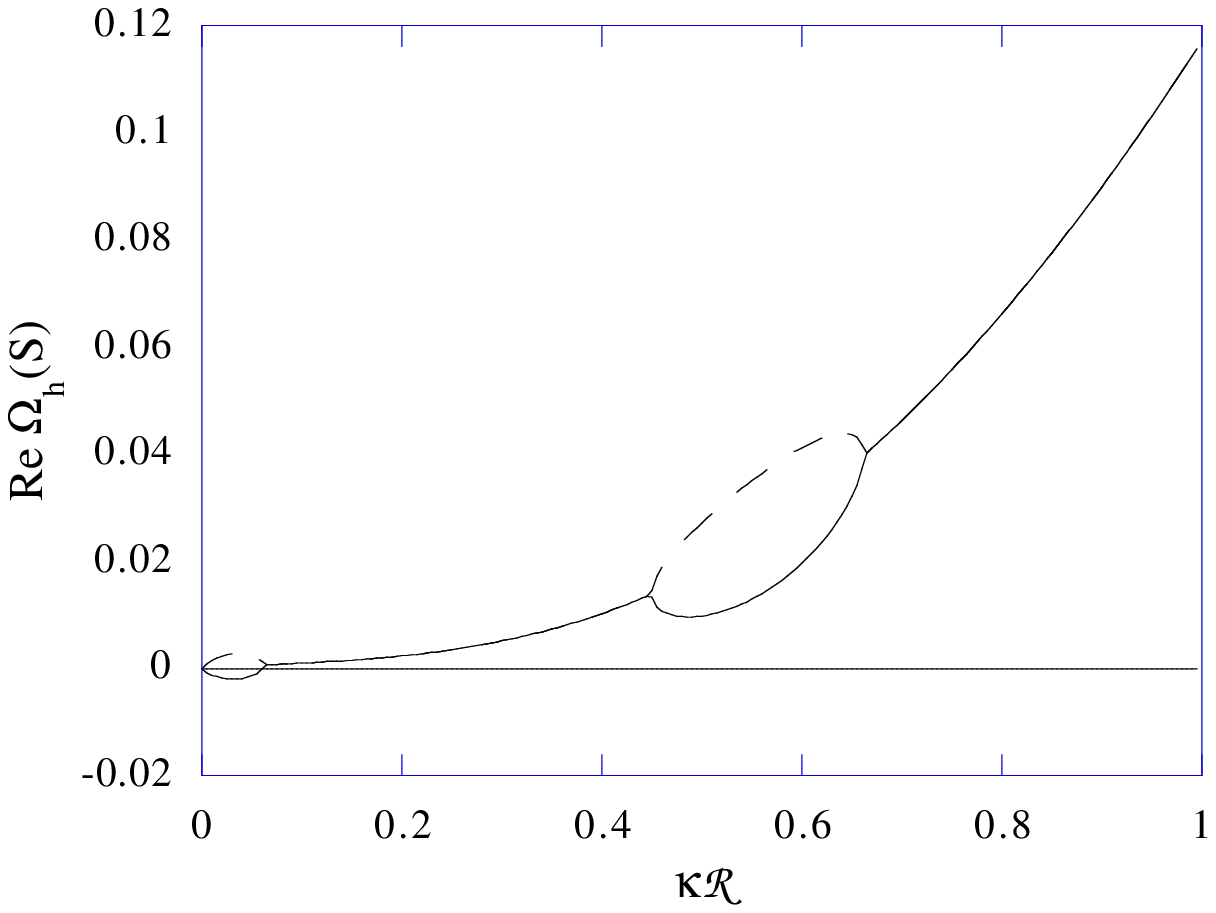}
\end{center}
\caption{\label{mup25gp16} Stability eigenvalues $Re[\Omega_{0,h}(S)]$ 
(solid line) and $Re[\Omega_{1,h}(S)]$ (dashed line) for $\mu=.25$ at $S = 
.16$.}
\end{figure}

Next we explore the effect of the viscosity ratio between the two 
phases, $\mu$.  When the viscosities of the two phases are equal, 
above the critical shear rate the cylinder is stable against 
perturbations at all wavelengths as we see in Fig.\ \ref{mu1gc}.  This 
is not the case for all $\mu$.  For general $\mu$, the shear flow does 
stabilize the varicose mode around the main instability at 
$\kappa_{max}$, but for some $\mu$ there is a residual instability 
left at small wave numbers (long wavelengths).  An example for $\mu = 
.25$ is shown in Figs.\ \ref{mup25gp14} and \ref{mup25gp16}.  Fig.\ 
\ref{mup25gp14} shows the lowest two stability eigenvalues at a low 
shear rate, when the lowest mode $\omega_{0,h}(S)$ is still unstable 
near the original maximally unstable wave number $\kappa_{max} {\cal 
R}\simeq .59$.  We see that there is an additional unstable region at 
small $\kappa$, separate from the main instability.  This long 
wavelength instability remains after the main instability has been 
stabilized by the shear flow, as shown in Fig.\ \ref{mup25gp16}.  The 
physical significance of this residual instability will be discussed 
further in Sec.\ \ref{relexp}; it does mean that the cylinder is still 
unstable to very long wavelength perturbations.  No residual 
instability was found for $.8 \lesssim \mu \lesssim 1.0$; as $\mu$ is 
either increased or decreased away from this range, a small 
instability appears smoothly from $\kappa=0$ and extends over 
increasingly large $\kappa$ as $\mu$ becomes correspondingly smaller 
than about $.8$ or larger than about $1.0$.

\begin{figure}
\begin{center}
\leavevmode
\epsfxsize= 4in
\epsfbox{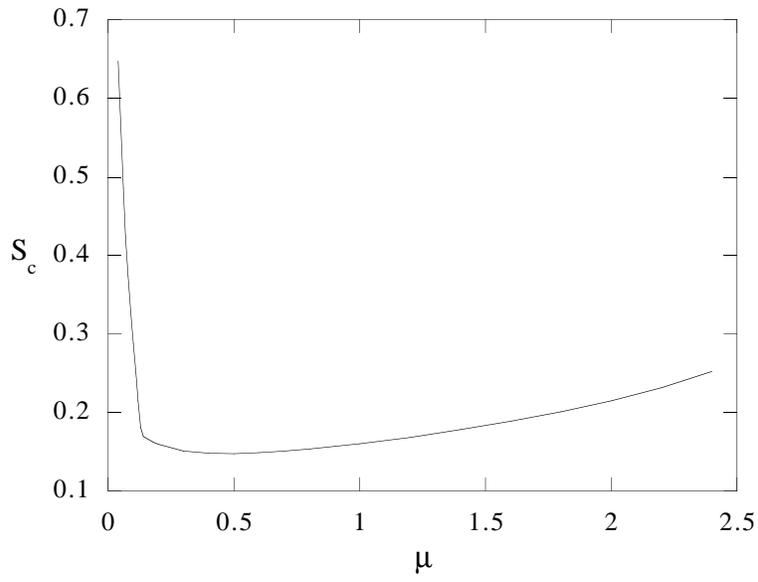}
\end{center}
\caption{\label{Scmu} Critical shear rate as a function of 
viscosity ratio $\mu$.}
\end{figure}

Nevertheless we can calculate the critical shear rate required to 
stabilize the original maximally unstable mode, as a function of 
$\mu$.  The result is shown in Fig.\ \ref{Scmu}.  The graph only 
includes values in the range $.04 \leq \mu \leq 2.4$ since for values 
of $\mu$ outside this range, the critical shear rate $S_{c}$ becomes 
larger than the damping rate of the $j=2$ mode; to extend the range of 
$\mu$ would therefore require including the $j=3$ mode and higher as 
$S_{c}$ increased.  [From Eq.\ (\ref{wk0}), we find 
$\Omega_{2,h}(k=0,\mu=.04) = .65$, $\Omega_{3,h}(k=0,\mu=.04) = .98$, 
$\Omega_{2,h}(k=0,\mu=2.4) = .20$, and $\Omega_{3,h}(k=0,\mu=2.4) = 
.29$ so the range in Fig.\ \ref{Scmu} is reasonable.]  $S_{c}(\mu)$ 
has a minimum near $\mu = .5$ and rises on either side, so that as the 
domain becomes either more or less viscous than about half the outside 
viscosity, it requires a higher shear rate to stabilize it.  The 
rather sharp bend near $\mu=.1$ is due to the fact that the maximally 
unstable varicose mode with growth rate 
$\Omega_{0,h}(k_{max}R,\mu,S=0)$ moves to lower wave numbers as $\mu$ 
decreases (see Fig.\ \ref{w0hmu}), while the $m=1$ mode in particular 
changes less with $\mu$ (Fig.\ \ref{w1hmu}).  The magnitudes of the 
two eigenvalues both increase as the inner viscosity decreases, but 
the growth rate of the $m=0$ mode does so more quickly and with a 
larger change in the dependence on $kR$.  Thus as $\mu$ decreases the 
interaction between these two modes changes in such a way as to result 
in the fairly sharp increase in shear rate necessary for stabilization 
for $\mu<.1$.

\subsection{Stabilization of the thermodynamic instability}

\begin{figure}
\begin{center}
\leavevmode
\epsfxsize= 4in
\epsfbox{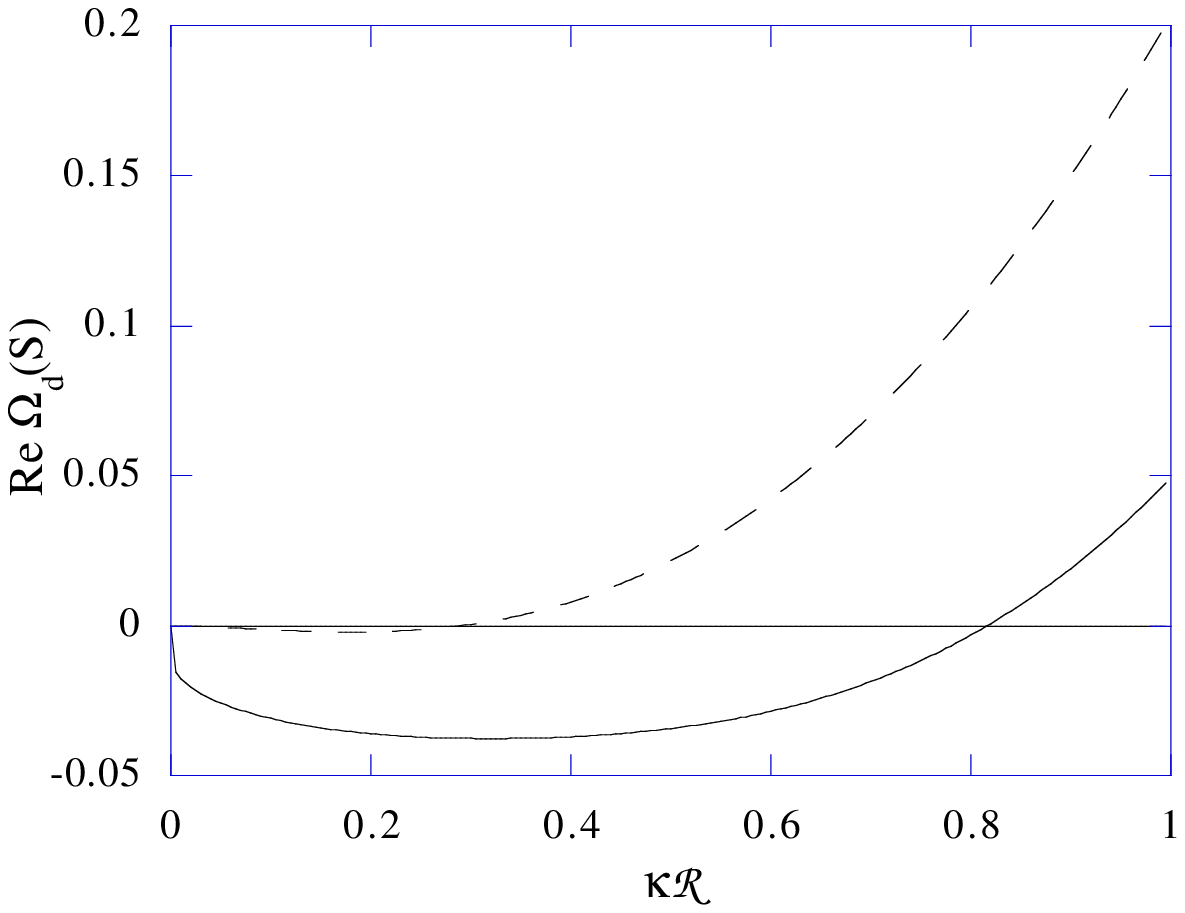}
\end{center}
\caption{\label{wdgp1} Real part of the dimensionless stability 
eigenvalues $\Omega_{0,d}(S)$ (solid curve) and $\Omega_{1,d}(S)$ 
(dashed curve) at $S^{*}=.1$ and $\mu=1$.}
\end{figure}

\begin{figure}
\begin{center}
\leavevmode
\epsfxsize= 4in
\epsfbox{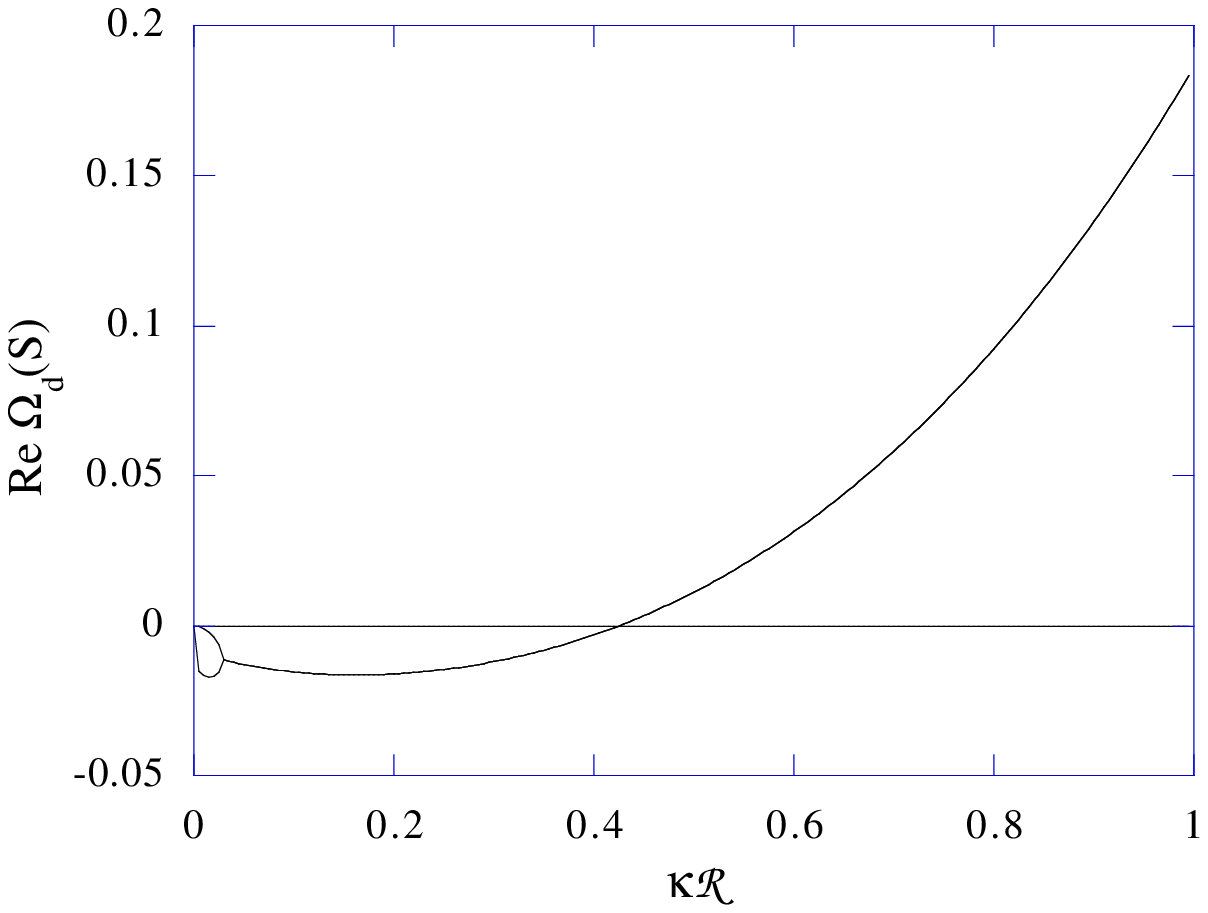}
\end{center}
\caption{\label{wdgp4} Real part of the dimensionless stability 
eigenvalues $\Omega_{0,d}(S)$ (solid curve) and $\Omega_{1,d}(S)$ 
(dashed curve) at $S^{*}=.4$ and $\mu=1$.}
\end{figure}

\begin{figure}
\begin{center}
\leavevmode
\epsfxsize= 4in
\epsfbox{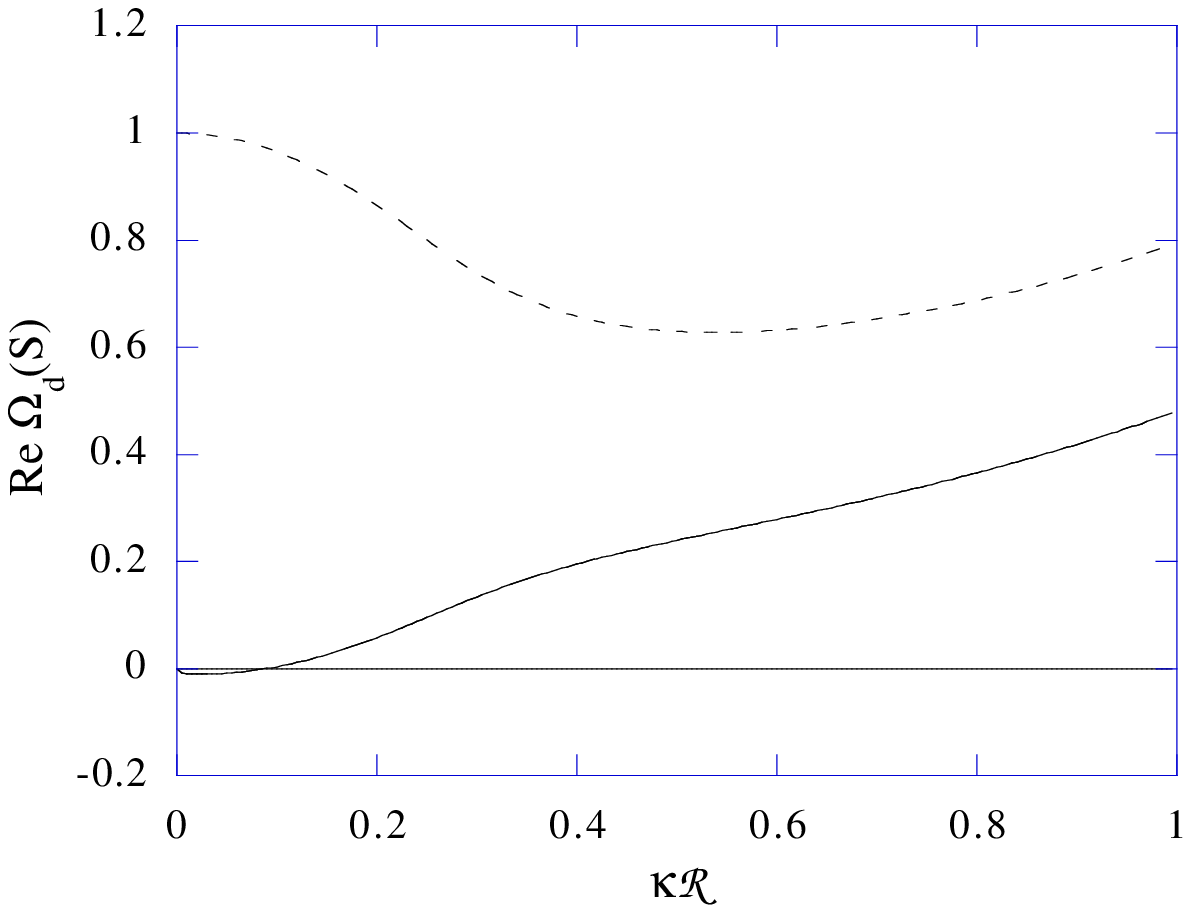}
\end{center}
\caption{\label{wdg2} Real part of the dimensionless stability 
eigenvalues $\Omega_{0,d}(S)$ (solid curve), $\Omega_{1,d}(S)$ (dashed 
curve), and $\Omega_{2,d}(S)$ (dotted curve) at $S^{*}=2$ and 
$\mu=1$.}
\end{figure}

In the last section we saw that the hydrodynamic Rayleigh instability 
can be partially or completely stabilized by the shear flow, depending 
on the viscosity contrast between the two phases.  The opposite limit 
is to consider what happens when the fluids are so viscous that the 
hydrodynamic terms are negligible.  Then the diagonal elements in Eq.\ 
(\ref{matrix2}) are just $\omega_{m,d}$ and the off-diagonal elements 
are the ones from the imposed flow $u_{s}$.  Fig.\ \ref{wdgp1} and 
Fig.\ \ref{wdgp4} show the dimensionless parts of the first two modes, 
$\Omega_{j,d}(S) \equiv \omega_{j,d}(S) R^{3} = \varpi_{j,d}(S) 
\tau_{\xi} {\cal R}^{3}/\xi^{3}$, for two different shear rates at 
$\mu=1$.  Here I have defined $S^{*}\equiv SR^{3}$ so that the trivial 
dependence on $R$ can be factored out.  Again as $S^{*}$ is increased 
the window of wave numbers for which the varicose mode is unstable 
becomes smaller.  Also, the mode of maximum instability moves to lower 
$\kappa$.  Fig.\ \ref{wdg2} shows all three modes in shear for a 
rather high shear rate, $S^{*}=2$.  The varicose mode has been mostly 
stabilized, with a small residual instability at small wave numbers.  
Including the fourth mode, $j=3$, allows us to raise the shear rate up 
to some fraction of the magnitude of the $j=4$ mode, 
$\Omega_{4,d}=64$.  (Note that including the $j=3$ mode does not 
change the lower three modes e.g.\ in Fig.\ \ref{wdg2} at all, since 
it is not mixed with them at low shear rates.)  The instability at 
small wave numbers moves to smaller and smaller $\kappa$ as $S^{*}$ is 
increased above $2$ towards $64$, but seems pinned near $\kappa=0$ and 
never quite disappears.  The changes with $S^{*}$ occur more slowly as 
$S^{*}$ is increased.  

Thus when the diffusive terms dominate the behavior, there is no 
well-defined critical shear rate for stabilization at all wave numbers 
$\kappa$.  Furthermore, unlike the case of the Rayleigh instability 
analyzed above, the mode which is maximally unstable at $S^{*}=0$ does 
not cross the axis at some well-defined shear rate independently of 
the residual instability at small wave number; instead the maximally 
unstable mode just shifts with $S^{*}$ towards smaller $\kappa$.  We 
could however define a critical shear rate for stabilization at any 
given (small) wave number $\kappa_{c}$; in this case the ``critical'' 
shear rate for stabilization will scale simply as
\begin{equation}
	S_{c}(\kappa_{c}) \propto \frac{\mu+1}{R^{3}} \,,
\end{equation}
since $\mu$ only enters in the off-diagonal shear flow terms.  Once 
again the critical shear rate is a decreasing function of the radius 
$R$, so for a given shear rate small cylinders will be unstable and 
large ones will be stable for wave numbers satisfying $\kappa > 
\kappa_{c}$.

\subsection{Relation to experiments}
\label{relexp}

In general the stability of the cylindrical domain will be determined 
by both hydrodynamic and diffusive effects, depending on the system 
parameters.  In this section I will first consider the parameters 
relevant to Hashimoto {\em et al}.'s experimental system 
\cite{string95}.  They have studied phase separation under shear flow 
in a pseudobinary mixture of polybutadiene (PB) and polystyrene (PS) 
in a common solvent of dioctylphthalate (DOP).  They find a 
correlation length $\xi \approx O(1000) $ {\AA}, a surface tension on 
the order of $10^{-4}$ erg, a diffusion constant on the order of 
$10^{-10} {\rm cm}^{2}$/s, the viscosity of PB/DOP $\nu_{PB} \approx 
1.2$ poise, and the viscosity of PS/DOB $\nu_{PS} \approx .3$ poise 
\cite{hashimoto2}.  For comparison with my results I will take as an 
example a viscosity ratio between the two phases of $\mu = .25$ so the 
cylindrical domain consists of the less viscous phase.  For the 
possible range of values of $\eta^{o}$ in the experiment, $\eta^{o} 
\approx .06 \sim .2$, the hydrodynamic terms in the diagonal matrix 
elements $\omega_{m}$ are significantly larger than the diffusive 
terms at all reasonable values of $\cal R$.  This is not surprising; 
at the large length scale of the domains, ${\cal R} >> \xi$, we would 
not necessarily expect the diffusive terms to be important.  Thus in 
this case the results of Sec.\ \ref{Rstable} apply with negligible 
modification.  The critical shear rate for stabilization of the 
cylinder at most $\kappa$, for $\mu=.25$ and $\eta^{o}=.1$, is
\begin{equation}
	\dot{\gamma}_{c} \tau_{\xi} = \frac{1.54 }{R}.
\end{equation}
Since the theory only applies for $R \geq 3$ or so (for smaller $R$ 
Eq.\ (\ref{profileR}) is no longer a good approximation), this shear 
rate is in the weak-shear regime, $\dot{\gamma} \tau_{\xi} < 1$, and 
is significantly smaller than the shear rate necessary for formation 
of the string phase seen in the experiments.  It is thus consistent 
that the long cylindrical domains seen experimentally are stable, 
since they are seen at shear rates that are well above the shear rate 
required for stabilization.  

\begin{figure}
\begin{center}
\leavevmode
\epsfxsize= 4in
\epsfbox{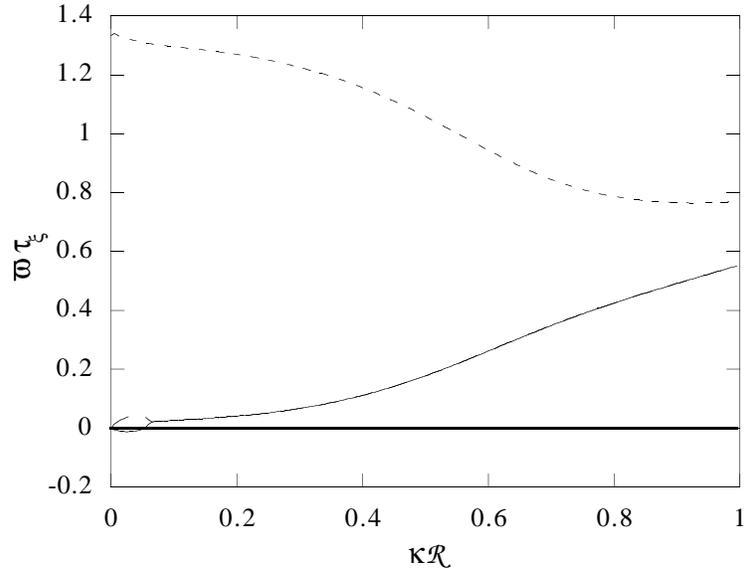}
\end{center}
\caption{\label{exp} Stability eigenvalues $Re[\varpi_{0}(S)]$ (solid line), 
$Re[\varpi_{1}(S)]$ (dashed line), and $Re[\varpi_{2}(S)]$ (dotted line) for 
$\dot{\gamma} \tau_{\xi}=1.5$, $\mu=.25$, $\eta^{o}=.1$ and ${\cal 
R}=4\xi$.}
\end{figure}

Fig.\ \ref{exp} shows the stability eigenvalues well above the 
critical shear rate, at $\dot{\gamma}\tau_{\xi}=1.5$.  Although this 
shear rate is in the strong-shear regime $\dot{\gamma} \tau_{\xi}>1$ 
for which the theory may not strictly be valid, it seems reasonable 
that the theory can be pushed into the strong shear regime at the 
large length scales $O({\cal R})$ considered here (if in the 
strong-shear condition $\tau_{\xi}$ is replaced by the typical time 
scale for domain fluctuations, then the theory is valid here).  For 
these particular parameter values, Hashimoto {\em et al}.\ found that 
the length of the strings seen in the experiment were on the order of 
$300 {\cal R}$.  This length could be explained by the residual 
instability at small wave numbers discussed in Sec.\ \ref{Rstable}.  
The wavelength of maximum instability in Fig.\ \ref{exp} is 
approximately $\lambda = 2\pi/\kappa \approx 250 {\cal R}$, so the 
length of the strings seen experimentally may be set by the residual 
long wavelength varicose instability in the shear flow.

\begin{figure}
\begin{center}
\leavevmode
\epsfxsize= 4in
\epsfbox{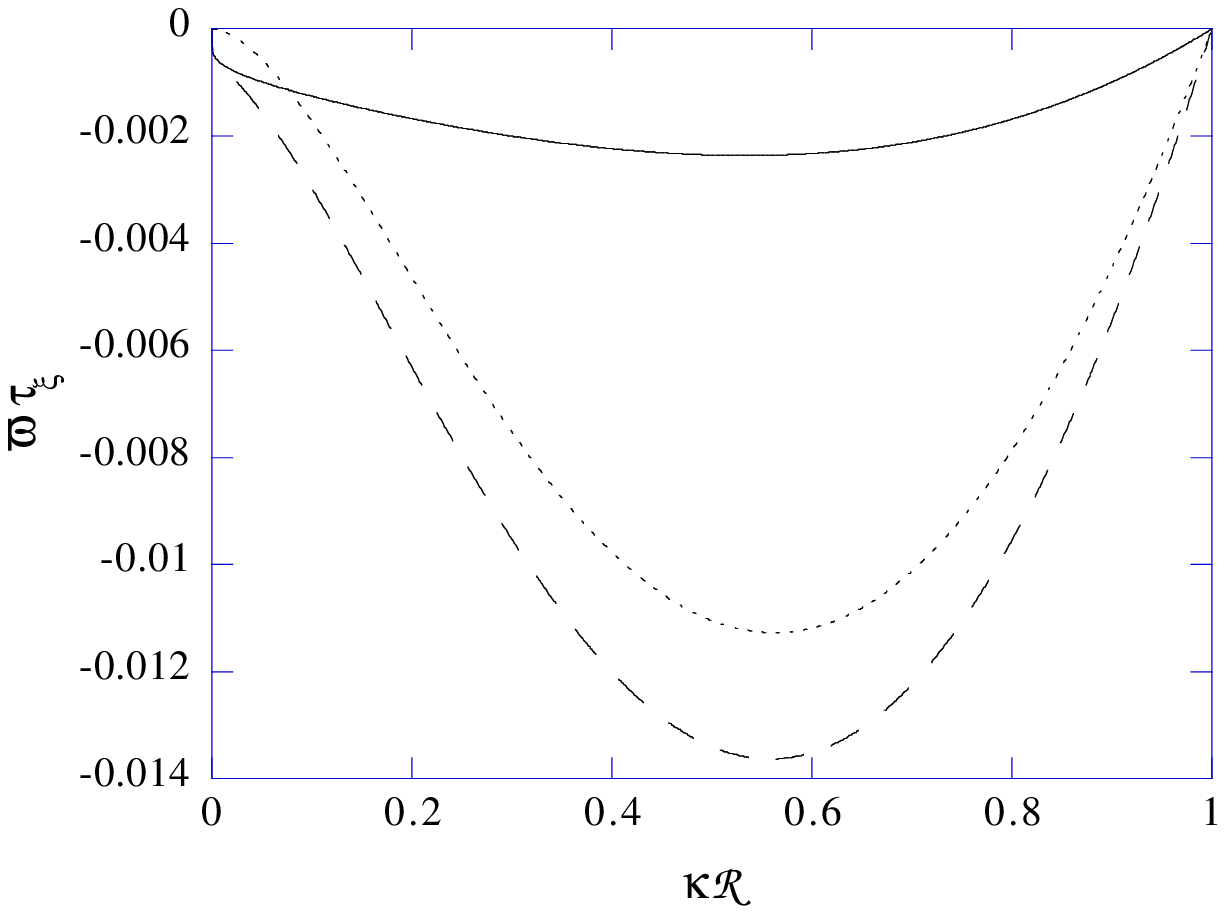}
\end{center}
\caption{\label{diff} Varicose mode: $\varpi_{0,d}$ (solid line), 
$\varpi_{0,h}$ (dotted line), and the total $\varpi_{0}$ (dashed line) 
for $\eta=.7$ and ${\cal R}=3\xi$ ($S=0$).}
\end{figure}

Next consider the case of near-critical binary fluids.  At the 
critical point, $\mu=1$ and $\eta$ is a universal number; for 
near-critical fluids it has the same order of magnitude, so we take 
the critical value $\eta \cong .7$.  For this value of $\eta$, the 
diffusive terms start to become noticeable at small radii, although 
they are still significantly smaller than the hydrodynamic terms at 
larger radii, e.g.\ $R>6$ or so.  Fig.\ \ref{diff} shows the varicose 
$m=0$ mode without the shear flow, for the diffusive and hydrodynamic 
terms separately at the smallest $R$ that is reasonable in the theory.  
The diffusive terms do change the magnitude of $\omega_{0}(S)$ and so 
will have a small quantitative effect on the results.  The 
stabilization by the shear flow is again very similar to the purely 
hydrodynamic case as illustrated in Figs.\  
(\ref{mu1gp1})--(\ref{mu1gp18}).  However, even for $\mu=1$ there is 
now once again a small residual instability at small $\kappa$ due to 
the diffusive part of the eigenvalues that persists at high shear, for 
the small radius ${\cal R}=3\xi$.  Note from Fig.\ \ref{diff} that at 
very small $\kappa$, $|\varpi_{0,d}| > |\varpi_{0,h}|$, so it is not 
surprising that the stability eigenvalues in shear resemble those of 
Fig.\ \ref{wdg2} at small $\kappa$.  Finally, the critical shear rate 
for stabilization of the main instability $S_{c}$ no longer scales 
exactly with $1/R$ at small $R$ due to the different scaling of the 
diffusive terms ($\omega_{d}\propto 1/R^{3}$), but the difference is 
small.  Thus, for near-critical binary fluids the effects of diffusive 
transport may be observable in string-like domains for sufficiently 
thin strings.

\section{Discussion}
\label{disc}

I have shown that shear flow can stabilize an isolated cylindrical 
domain in the two-phase state of a phase-separating binary fluid 
against varicose instabilities, by mixing the varicose mode with the 
other nonaxisymmetric perturbation modes of the cylinder.  
Essentially, the shear flow distorts the varicose mode by convecting 
one side of the cylindrical interface with respect to the other, 
eliminating the special axisymmetric, pinching character of the 
varicose mode that drove the instability.  Both the hydrodynamic 
Rayleigh instability and the thermodynamic, diffusive instability are 
suppressed by the shear flow, although there are residual 
instabilities at small wave numbers in the limit that the diffusive 
terms dominate and also for viscosity ratios $\mu$ outside the range 
$.8 \lesssim \mu \lesssim 1.0$.  Other authors have considered the 
effect of flows on the Rayleigh instability, but this is the first 
study focused on the effect of the nonaxisymmetric nature of shear 
flow on the Rayleigh instability.

Comparing with the experimental results of Hashimoto {\em et al}., I 
found the mechanism presented here for stabilization of the 
cylindrical domains is consistent with the stable ``string'' phase 
seen experimentally, and that the lengths of the strings may be set by 
the residual instabilty at long wavelengths.  However, it should be 
noted that the stability of a cylindrical domain in shear flow does 
not act as a criterion for the observed relationship between the shear 
rate and the radius of the domains seen in the string phase.  The 
domains in the string phase are formed through a dynamic process; the 
observed radius is not a parameter of the system (as in this 
calculation) but rather is determined through the self-organization 
process as the shear flow competes with coarsening in the 
phase-separating system.  I have merely demonstrated a mechanism by 
which these macroscopically long, cylindrical domains may be 
stabilized by the shear flow.  Although a few experiments have looked 
at the breakup of the string-like domains after complete cessation of 
the shear flow \cite{kim,takebe1,takebe2}, it would be interesting to 
do a careful experimental study of the shear rate at which the strings 
first begin to be unstable to see if the Rayleigh and/or thermodynamic 
varicose instabilities explored here are the main breakup mechanisms in 
these systems.  If so then the strings should be unstable below the 
critical shear rate $\dot{\gamma}_{c}$ found in this paper.

The results presented here may also shed light on why the shear flow 
can halt the phase separation and result in a dynamic steady state, 
even in the weak-shear regime, $\dot{\gamma} \tau_{\xi}<1$.  In a 
concentrated phase-separating fluid when the two phases are both 
percolated so that the domains form a connected bicontinuous pattern, 
the coarsening is dominated by curvature effects.  Qualitatively we 
can think of a piece of the interconnected structure as a cylinder of 
fluid immersed in the other phase.  This cylinder is susceptible to 
the varicose instabilities considered here, and particularly to the 
Rayleigh instability, which leads to breakup of the cylindrical region 
into spheres.  Siggia \cite{siggia} used this picture to explain the 
coarsening rates seen in concentrated binary fluids.  Since the shear 
flow suppresses these instabilities, one might expect that it could 
stabilize an anisotropic, bicontinuous morphology against further 
coarsening.  For a given shear rate, when the domains are relatively 
small they will be unstable and will coarsen, but once the typical 
length scale has grown to the critical radius for stabilization by the 
shear flow, ${\cal R}_{c}(\dot{\gamma})$, the parts of the 
bicontinuous structure that are cylindrical and aligned with the flow 
will no longer be unstable.  This then provides a mechanism for the 
creation of the nonequilibrium dynamic steady state seen in 
concentrated phase-separating fluids in shear flow.

\section*{Acknowledgments}

I would like to thank Jim Langer and Glenn Fredrickson for many 
helpful discussions and support.  I thank T. Hashimoto, A. Onuki, and 
E. Moses for useful discussions and correspondence, and J. Goveas for 
references to the appropriate fluid dynamics literature.  I would 
also like to thank the University of California, Santa Barbara for 
financial support.  This work was supported by the MRL Program of the 
National Science Foundation under Award No.\ DMR 96-32716 and by the 
U.S. DOE Grant No.\ DE-FG03-84ER45108.

\end{multicols}

\appendix

\section*{Calculation of the perturbed velocity field}

The perturbed velocity field ${\bf v}_{m}$ corresponding to a 
perturbation of the interface given by $r=R-\epsilon e^{im\theta + 
ikx}$ must satisfy the hydrodynamic equations
\begin{equation}
	\nabla^{2} {\bf v} = \frac{1}{\eta} {\bf \nabla}P,
	\label{hydro3.2}
\end{equation}
\begin{equation}
	0 = {\bf \nabla} \cdot {\bf v},
	\label{incomp3.2}
\end{equation}
\begin{equation}
	\nabla^{2} P = 0.
	\label{peq.2}
\end{equation}
To solve for $\bf v$ we again follow Happel and Brenner \cite{happel}.  
The velocity is expanded as in Eq.\ (\ref{vnm}),
\begin{displaymath}
	{\bf v}_{m}(r,\theta) = \sum_{n} e^{in\theta}{\bf v}_{nm}(r).
\end{displaymath}
We will solve for each coefficient ${\bf v}_{nm}$ separately so we 
take ${\bf v} \propto e^{in\theta}$.  We start by solving Eq.\ 
(\ref{peq.2}) for the pressure.  Let $P = P_{n}(r) e^{in\theta + 
ikx}$; then in general
\begin{equation}
	P_{n}(r) = \left\{ \begin{array}{ll}
	p_{1} I_{n}(kr) & r < R \\
	p_{2} K_{n}(kr) & r > R, 
	\end{array} \right. 
	\label{pressure2}
\end{equation}
where $p_{1}$ and $p_{2}$ are constants.  The perturbed velocity then 
satisfies the inhomogeneous Laplace equation (\ref{hydro3.2}), so the 
solution will consist of a general solution to the homogeneous part 
plus a particular solution, ${\bf v} = {\bf v}^{p}+{\bf v}^{g}$.  
Writing $v_{x} = i v_{x,n}(r) e^{in\theta + ikx}$, the $x$ component 
of Eq.\ (\ref{hydro3.2}) is
\begin{equation}
	iv_{x,n}'' + \frac{1}{r}i v_{x,n}' - \left(k^{2} + 
	\frac{n^{2}}{r^{2}}\right) 
	iv_{x,n} = \frac{ik}{\eta} P_{n}(kr).
\end{equation}
A particular solution is given by 
\begin{equation}
	v^{p}_{x,n} = \left \{ \begin{array}{ll}
	i c^{p}_{1} I_{n}(kr) + \frac{i\pi_{1}}{2} kr I_{n}'(kr) & r < R \\
	i c^{p}_{2} K_{n}(kr) +  \frac{i\pi_{2}}{2} kr K_{n}'(kr) & 
	r > R, \end{array} \right. 
\end{equation}
where $\pi_{1} \equiv p_{1}/\eta^{i} k$ and $\pi_{2} \equiv 
p_{2}/\eta^{o} k$.  The $r$ and $\theta$ components of $v$ satisfy 
Eq.\ (\ref{vcoupled}) without the extra $\phi$ term; assuming they are 
expanded as in Eq.\ (\ref{vexp}) the particular solutions to Eq.\ 
(\ref{hydro3.2}) are
\begin{equation}
	v^{p}_{r,n} + i v^{p}_{\theta,n} =  \left \{ \begin{array}{ll}
    \pi_{1} kr I_{n+1}'(kr) & r< R \\
	\pi_{2} kr K_{n+1}'(kr) & r<R, \end{array} \right. 
\end{equation}
\begin{equation}
	v^{p}_{r,n} - i v^{p}_{\theta,n} =  \left \{ \begin{array}{ll}
	\pi_{1} kr I_{n-1}'(kr) & r< R \\
	\pi_{2} kr K_{n-1}'(kr) & r<R
	\end{array} \right. 
\end{equation}
(the solutions to the homogeneous part of the equation will be 
included in the general solution below and so are not needed here).  
The components of ${\bf v}^{p}$ must satisfy the continuity equation 
(\ref{incomp3.2}):
\begin{displaymath}
	\frac{dv^{p}_{r,n}}{dr} + \frac{v^{p}_{r,n}}{r} + \frac{in}{r} 
	v^{p}_{\theta,n} + ikv^{p}_{x,n} = 0.
\end{displaymath}
This determines the constants $c^{p}_{1}=\pi_{1}/2$ and 
$c^{p}_{2}=\pi_{2}/2$.

Next we need a general solution to the homogeneous part of Eq.\ 
(\ref{hydro3.2}). Again these are just the appropriate solutions of 
the Laplace equation:
\begin{equation}
	v^{g}_{x,n} = \left \{ \begin{array}{ll}
	i c_{1} I_{n}(kr)  & r < R \\
	i c_{2} K_{n}(kr)  & r > R, \end{array} \right. 
\end{equation}
\begin{equation}
	v^{g}_{r,n} + i v^{g}_{\theta,n} =  \left \{ \begin{array}{ll}
	a^{+} I_{n+1}(kr)  & r< R \\
	b^{+} K_{n+1}(kr) & r>R, \end{array} \right. 
\end{equation}
\begin{equation}
	v^{g}_{r,n} - i v^{g}_{\theta,n} =  \left \{ \begin{array}{ll}
	a^{-} I_{n-1}(kr)  & r< R \\
	b^{-} K_{n-1}(kr)  & r>R, \end{array} \right. 
\end{equation}
and once again enforcing incompressibility gives 
$c_{1} = (a^{+}+a^{-})/2$ and $c_{2} = -(b^{+}+b^{-})/2$. The 
result is a solution for ${\bf v}_{n}$ containing six unknown constants, 
$a^{\pm}$, $b^{\pm}$, and $\pi_{1,2}$:
\begin{mathletters}
\label{vg}
\begin{equation}
	v_{x,n} = \frac{i}{2} \left \{ \begin{array}{ll}
	 (a^{+}+a^{-}) I_{n}(kr) + \pi_{1} \left[I_{n}(kr) + kr 
	 I_{n}'(kr)\right] & r < R \\
	-(b^{+}+b^{-}) K_{n}(kr) +  \pi_{2}\left[K_{n}(kr) + kr 
	K_{n}'(kr)\right] & r > R, \end{array}\right.  
\end{equation}
\begin{equation}
	v_{r,n}  = \frac{1}{2} \left \{ \begin{array}{ll}
	a^{+} I_{n+1}(kr) +	a^{-} I_{n-1}(kr) + \pi_{1} kr 
	\left[I_{n+1}'(kr)+I_{n-1}'(kr)\right] & r< R \\
	b^{+} K_{n+1}(kr) + b^{-} K_{n-1}(kr) - \pi_{2} kr\left[K_{n+1}'(kr) + 
	K_{n-1}'(kr)\right] & r>R, 
	\end{array} \right. 
\end{equation}
\begin{equation}
	v_{\theta,n} = \frac{-i}{2} \left \{ \begin{array}{ll}
	a^{+} I_{n+1}(kr) -	a^{-} I_{n-1}(kr) + \pi_{1} kr 
	\left[I_{n+1}'(kr)-I_{n-1}'(kr)\right] & r< R \\
	b^{+} K_{n+1}(kr) - b^{-} K_{n-1}(kr) - \pi_{2} kr\left[K_{n+1}'(kr) - 
	K_{n-1}'(kr)\right] & r>R.
	\end{array}  \right. 
\end{equation}
\end{mathletters}
These equations give the general solutions for the coefficients ${\bf 
v}_{nm}$ in Eq.\ (\ref{vnm}).

It now remains to apply the appropriate boundary conditions at the 
interface to specify the remaining constants.  These boundary 
conditions will apply to the total velocity and stress fields.  
Letting $\epsilon$ denote the amplitude of the small perturbations, 
recall we have
\begin{eqnarray*}
	{\bf u} &  = & {\bf u}_{s} + \epsilon {\bf v}, \\
	P & = & P_{s} + \epsilon P, \\
	{\bf \Pi} &  = & {\bf \Pi}^{s} + \epsilon \mbox{\boldmath$\sigma$},
\end{eqnarray*}
where $P_{s}$ is the steady state pressure, ${\bf \Pi}^{s}$ the steady 
state stress tensor, and \boldmath $\sigma$ \unboldmath the perturbed 
stress tensor.  From Eq.\ (\ref{us}), the components of the steady 
state stress tensor are
\begin{equation}
	\Pi_{xr}^{s}  =  \frac{2S
	\eta^{i}}{\mu+1} \cos \theta,
\end{equation}
\begin{equation}
	\Pi_{x\theta}^{s} = \left\{ \begin{array}{ll}
	-  \frac{\textstyle 2S\eta^{i}}{\textstyle \mu+1} \sin \theta & r < R \\
	-\eta^{o} S \sin \theta - \frac{\textstyle \eta^{o}}{\textstyle r^{2}}\left( 
	\frac{\textstyle 2SR^{2}}{\textstyle \mu+1} - SR^{2} \right) \sin 
	\theta & r>R \end{array} \right. 
	\label{pixtheta}
\end{equation} 
(note ${\bf \Pi}$ is symmetric).  We see that $\Pi^{s}_{xr}$ is 
continuous across the interface but $\Pi^{s}_{x\theta}$ has a jump 
across the interface when $\eta^{i} \neq \eta^{o}$.  The difference in 
the steady state pressure $P_{s}$ across the interface is simply the 
Laplace pressure across a cylindrical interface $\sigma/{\cal R}$, 
which in our dimensionless variables (the dimensionless surface 
tension is $\bar{\sigma} = 2/3$) is
\begin{equation}
	P_{s}^{i} - P_{s}^{o} = \frac{2}{3R}.
	\label{Laplace}
\end{equation}
The boundary conditions are \cite{joseph}:
\begin{itemize}
\item Continuity of the velocity across the interface, ${\bf u}^{i} = 
{\bf u}^{o}$ 
\item Continuity of the tangential stress across the interface, $\Pi^{i}_{t} = 
\Pi^{o}_{t}$ 
\item Jump in the normal stress across the interface due 
to the mean curvature $H$, $\Pi^{i}_{n} - \Pi^{o}_{n} = 4H/3$ (here 
$H$ is dimensionless)
\end{itemize}

To apply these boundary conditions we need to evaluate the appropriate 
components of the stress tensor on the deformed interface 
\cite{mikami}.  The location of the cylindrical interface for mode $m$ 
is $r = R - \epsilon e^{im\theta + ikx}$ so the unit normal to the 
interface is
\begin{equation}
	{\bf \hat{n}} = {\bf \hat{r}} + \frac{im}{R} \epsilon e^{im\theta + ikx} 
	{\bf \hat{\theta}} + ik\epsilon e^{im\theta + ikx} {\bf \hat{x}} + 
	O(\epsilon^{2}) \,.
\end{equation}
The hydrodynamic force on the perturbed surface is ${\bf F} = {\bf \Pi} 
\cdot {\bf \hat{n}}$. To lowest order in $\epsilon$, the three components 
of ${\bf F}$ are
\begin{eqnarray}
	F_{r} & = & -P_{s} + \epsilon \sigma_{rr} + ik\epsilon e^{im\theta + 
	ikx} \Pi_{rx}^{s}, \\
	F_{\theta} & = & \epsilon \sigma_{\theta r} - \frac{im}{R}\epsilon 
	e^{im\theta + ikx} P_{s} + ik\epsilon e^{im\theta + ikx} 
	\Pi_{\theta x}^{s}, \\
	F_{x} & = & \Pi_{xr}^{s} + \epsilon \sigma_{xr} + \frac{im}{R} \epsilon 
	e^{im\theta + ikx} \Pi_{x\theta}^{s} 
	 - ik\epsilon e^{im\theta + ikx} P_{s}.
\end{eqnarray}
The normal stress is simply 
\begin{equation}
	F_{n} = {\bf F} \cdot {\bf \hat{n}} = -P_{s} + \epsilon \sigma_{rr} + 
	2ik\epsilon e^{im\theta + ikx} \Pi_{rx}^{s} + O(\epsilon^{2}).
	\label{norm}
\end{equation}
The two tangential components of the stress can be found from ${\bf 
F}_{t} = {\bf F} - F_{n} {\bf \hat{n}}$, giving
\begin{eqnarray}
	(F_{t})_{\theta} & = & \epsilon \sigma_{\theta r} + ik\epsilon e^{im\theta + 
	ikx} \Pi_{\theta x}^{s} + O(\epsilon^{2}), \\
	(F_{t})_{x} & = & \Pi_{xr}^{s} + \epsilon \sigma_{xr} + \frac{im}{R} \epsilon 
	e^{im\theta + ikx} \Pi_{x\theta}^{s} +  O(\epsilon^{2}).
	\label{tang}
\end{eqnarray}
Finally, the mean curvature $H$ is
\begin{equation}
	H = -\frac{1}{2R} + \epsilon \frac{k^{2}R^{2} + m^{2} -1}{2R^{2}}  
	e^{im\theta + ikx} + O(\epsilon^{2});
\end{equation}
the first term is the steady state pressure difference given by Eq.\ 
(\ref{Laplace}).  Since the stationary velocity ${\bf u}_{s}$ already 
satisfies the boundary conditions, the perturbed velocity ${\bf v}$ 
must satisfy them separately.  Denote the difference between 
quantities inside and outside the cylinder at $r=R$ by $[[f]] = 
f^{i}-f^{o}$.  Then keeping in mind Eq.\ (\ref{Laplace}) and that 
$\Pi_{xr}^{s}$ is continuous across the interface, from Eqs.\ 
(\ref{norm})--(\ref{tang}) the six boundary conditions become:
\begin{mathletters}
\label{bcs}
\begin{eqnarray}
	\left[ \left[ v_{r} \right]\right] & = & \left[ \left[ v_{\theta} 
	\right]\right] = \left[ \left[ v_{x}\right]\right] = 0, \\
	 \left[ \left[ \sigma_{rr}\right]\right] & = & \frac{2}{3} 
	 \frac{k^{2}R^{2}+m^{2}-1}{R^{2}} e^{im\theta + ikx}, \\
	\left[ \left[ \sigma_{\theta r}\right]\right] 
	 & = & -\left[ \left[ ik e^{im\theta + ikx} \Pi^{s}_{\theta x} 
	 \right] \right], \\
	 \left[ \left[ \sigma_{xr} \right]\right] & = & -\left[ \left[ \frac{im}{R} 
	 e^{im\theta + ikx} \Pi^{s}_{\theta x} \right]\right].
\end{eqnarray}
\end{mathletters}
Since $\Pi^{s}_{\theta x}$ depends on $\theta$, for a given 
perturbation mode $m$ of the cylinder the solution ${\bf 
v}_{m}(r,\theta)$ given by Eq.\ (\ref{vnm}) will contain more than one 
coefficient ${\bf v}_{nm}$.  The perturbed stress tensor \boldmath ${\bf 
\sigma}$ \unboldmath depends only on ${\bf v}$ so we can write
\begin{equation}
	\mbox{\boldmath $\sigma$}(r,\theta,x) = \sum_{n} \mbox{\boldmath 
	$\sigma$}_{n}(r) e^{in\theta + ikx}.
	\label{stotal}
\end{equation}
Noting that $\sin \theta = (e^{i\theta} - e^{-i\theta})/2i$ and using 
Eq.\ (\ref{pixtheta}) at $r=R$ we can write
\begin{equation}
	\left[\left[i e^{im\theta + ikx} \Pi_{\theta x}^{s}\right]\right] 
	= \left(-\frac{S \eta^{i}}{\mu+1} + \frac{S \eta^{o}}{\mu+1} 
	\right) \left(e^{i(m+1)\theta} - e^{i(m-1)\theta}\right).
\end{equation}
Substituting in the full sums in Eq.\ (\ref{vnm}) and Eq.\ 
(\ref{stotal}) and matching terms with the same $\theta$ dependence in 
Eqs.\ (\ref{bcs}) then gives for the boundary conditions on each 
coefficient ${\bf v}_{nm}$,
\begin{mathletters}
\label{bcs2}
\begin{eqnarray}
	\left[\left[ v_{r,nm} \right]\right] & = & \left[\left[ 
	v_{\theta,nm} \right]\right] = \left[\left[ v_{x,nm} 
	\right]\right] = 0, \\
	\left[\left[ \sigma_{rr,nm} \right]\right] & = & \frac{2}{3} 
	\frac{k^{2}R^{2}+m^{2}-1}{R^{2}} \delta_{mn}, \label{normal} \\
	\left[\left[ \sigma_{\theta r,nm} \right]\right]
	 & = & \left(\eta^{i}-\eta^{o}\right) \frac{kS}{\mu+1} 
	 \left(\delta_{n=m+1} - \delta_{n=m-1}\right), \label{tang1} \\
	\left[\left[ \sigma_{xr,nm} \right]\right] 
	& = & \left(\eta^{i}-\eta^{o}\right) \frac{mS}{R(\mu+1)} 
	\left(\delta_{n=m+1} - \delta_{n=m-1}\right).  \label{tang2}
\end{eqnarray}
\end{mathletters}
The right-hand side of the normal stress condition Eq.\ (\ref{normal}) 
is only nonzero for $n=m$, and the right-hand sides of the tangential 
stress conditions Eqs.\ (\ref{tang1}) and (\ref{tang2}) are only 
nonzero for $n=m \pm 1$.  This immediately shows that for a given 
perturbation mode $m$ of the interface, the only nonzero coefficients 
$v_{nm}$ will be those for which $n=m$, $n=m+1$, and $n=m-1$.  The 
eigenvalue equation (\ref{matrix1}) is thus tridiagonal.  To solve for 
each matrix element $v_{r,nm}(r=R)$ (see Sec.\ \ref{vmu}), we just 
substitute the appropriate general solution from Eqs.\ (\ref{vg}) into 
the boundary conditions (\ref{bcs2}) and solve the resulting system of 
algebraic equations for the unknown constants.  This is best done 
numerically given the algebra involved and was solved using a standard 
algorithm \cite{NR}.

We can however see analytically how the matrix elements depend on 
$\eta^{o}$ and $R$.  First consider the boundary condition on the 
normal stress, Eq.\ (\ref{normal}).  Writing out the stress tensor we 
have
\begin{eqnarray}
	\left[\left[\sigma_{rr,m}\right]\right] & = & \left[\left[-p_{m} + 2\eta 
	\frac{\partial v_{r,m}}{\partial r}\right]\right] \nonumber \\
	 & = & -p_{1}I_{m}(kR) + 2\eta^{i}k v_{r,m}^{\prime ,i}(kR) + 
	 p_{2}K_{m}(kR) - 2\eta^{o}k v_{r,m}^{\prime ,o}(kR) \nonumber \\
	  & = & \frac{2}{3} \frac{k^{2}R^{2}+m^{2}-1}{R^{2}},
	  \label{norm2}
\end{eqnarray}
using Eq.\ (\ref{pressure2}) for the pressure.  The primes indicate 
differentiation with respect to the argument of $v_{r}$, $q=kR$.  
Dividing both sides by $\eta^{o}k$ leaves
\begin{equation}
	-\pi_{1} \mu I_{m}(q) + 2\mu v_{r,m}^{\prime,i}(q) + \pi_{2} K_{m}(q) - 
	2v_{r,m}^{\prime ,o}(q) = \frac{2}{3\eta^{o}R} \frac{q^{2}+m^{2}-1}{q}.
\end{equation}
The left-hand side now depends only on the six integration constants and 
on the dimensionless parameters $q$ and $\mu$ (which remain 
dimensionless when written in the original variables).  For the 
tangential stress in Eq.\ (\ref{tang1}),
\begin{eqnarray*}
	\left[\left[\sigma_{\theta r,nm} \right]\right] & = & \left[\left[ 
	\eta \left(\frac{1}{R} \frac{\partial v_{r}}{\partial \theta}+ 
	\frac{\partial v_{\theta}}{\partial r} - \frac{v_{\theta}}{R} 
	\right) \right]\right] \nonumber \\
	 & = & \left[\left[ \eta \left(\frac{im}{R} v_{r,n}(kR) + k 
	 v_{\theta,n}'(kR) - \frac{1}{R} v_{theta,n}(kR) \right) 
	 \right]\right] \nonumber \\
	 & = & \left(\eta^{i}-\eta^{o}\right) \frac{kS}{\mu+1} 
	 \left(\delta_{n=m+1} - \delta_{n=m-1}\right).
\end{eqnarray*}
Dividing both sides by $\eta^{o}$ and multiplying by $R$ gives
\begin{equation}
	 \mu \left[imv_{r,n}^{i}(q) + qv_{\theta,n}^{\prime ,i}(q) - 
	v_{\theta,n}^{i}(q)\right] - imv_{r,n}^{o}(q) - 
	qv_{\theta,n}^{\prime ,o}(q) + 
	v_{\theta,n}^{o}(q) 
	=  \frac{\mu - 1}{\mu + 1} qS \left(\delta_{n=m+1} 
	- \delta_{n=m-1}\right).
	\label{tang1.2}
\end{equation}
Similarly, from Eq.\ (\ref{tang2}) we have
\begin{eqnarray*}
	\left[\left[\sigma_{xr,nm}\right]\right] & = & \left[\left[ \eta 
	\left( \frac{\partial v_{x}}{\partial r} + \frac{\partial 
	v_{r}}{\partial x} \right) \right]\right] \nonumber \\
     & = & \left[\left[ \eta \left(k v_{x,n}'(kR) + ik v_{r,n}(kR) 
     \right)\right]\right] \nonumber  \\
     & = & \left(\eta^{i}-\eta^{o}\right) \frac{mS}{R(\mu+1)} 
	 \left(\delta_{n=m+1} - \delta_{n=m-1}\right).
\end{eqnarray*}
Dividing by $\eta^{o}k$ gives
\begin{equation}
	\mu \left[ v_{x,n}^{\prime ,i}(q) + iv_{r,n}^{i}(q) \right] - 
	v_{x,n}^{\prime ,o}(q) - iv_{r,n}^{o}(q) = \frac{\mu - 1}{\mu +1} 
	\frac{mS}{q} \left(\delta_{n=m+1} - \delta_{n=m-1}\right).
	\label{tang2.2}
\end{equation}
Both Eq.\ (\ref{tang1.2}) and Eq.\ (\ref{tang2.2}) depend only on 
the integration constants, $q$, $\mu$, and $S$.  In calculating 
the diagonal velocity matrix elements $v_{r,mm}(R)$, all 
right-hand sides are zero except in the normal stress equation 
(\ref{norm2}), so we see that in this case the integration constants and thus 
the diagonal elements as well will scale as $1/\eta^{o}R$.  For 
the off-diagonal elements the only nonzero right-hand sides are 
from the tangential stress conditions, so the off-diagonal elements 
will only depend on $q$, $\mu$ and $S$, and not on $R$ or the 
magnitude of the viscosities.  In the absence of the shear flow 
this implies that the stability eigenvalues coming from the 
hydrodynamic terms scale as $1/\eta^{o}R$.

Finally, the equations (\ref{bcs2}) can be solved analytically at 
$k=0$.  In this case the general solutions to the modified Bessel 
equation become $r^{n}$ and $r^{-n}$, and also the left-hand sides of 
Eqs.\ (\ref{tang1}) and (\ref{tang2}) are zero at $k=0$.  This 
simplifies the algebra considerably.  Following the same procedure as 
outlined above, we find that the velocity matrix elements and 
therefore the stability eigenvalues at $k=0$ are diagonal and 
independent of the shear rate $S$:
\begin{equation}
	\omega_{m,h}(k=0) = \frac{m}{3\eta^{o} R (\mu +1)}, \quad m \geq 2,
	\label{k0exact}
\end{equation}
with $\omega_{0,h}(k=0) = \omega_{1,h}(k=0) = 0$.  

\begin{multicols}{2}

\end{multicols}

\newpage

\begin{thebibliography}{10}
\bibitem{note1}
There seems to be some debate in the literature over whether the 
appropriate time scale to use in dilineating weak vs. strong-shear 
regimes in two-phase states should be $\tau_{\xi}$  or the time scale 
associated with the growth/fluctuations of the domains.
\end{thebibliography}
\end{document}